\documentclass[superscriptaddress,aps,prc,showkeys,showpacs,twocolumn,10pt]{revtex4}
\usepackage[OT1,T1]{fontenc}
\usepackage{graphicx}
\usepackage{rotating}
\begin{document}

\title{Measurement of the $pn \to pp\pi^0\pi^-$ Reaction in Search for the
  Recently Observed Resonance Structure in $d\pi^0\pi^0$ and $d\pi^+\pi^-$
  systems}  
\date{\today}


\newcommand*{\IKPUU}{Division of Nuclear Physics, Department of Physics and 
 Astronomy, Uppsala University, Box 516, 75120 Uppsala, Sweden}
\newcommand*{\ASWarsN}{Department of Nuclear Physics, National Centre for 
 Nuclear Research, ul.\ Hoza~69, 00-681, Warsaw, Poland}
\newcommand*{\IPJ}{Institute of Physics, Jagiellonian University, ul.\ 
 Reymonta~4, 30-059 Krak\'{o}w, Poland}
\newcommand*{\PITue}{Physikalisches Institut, Eberhard--Karls--Universit\"at 
 T\"ubingen, Auf der Morgenstelle~14, 72076 T\"ubingen, Germany}
\newcommand*{\Kepler}{Kepler Center for Astro and Particle Physics, Eberhard 
 Karls Universit\"at T\"ubingen, Auf der Morgenstelle~14, 72076 T\"ubingen, 
 Germany}
\newcommand*{\MS}{Institut f\"ur Kernphysik, Westf\"alische 
 Wilhelms--Universit\"at M\"unster, Wilhelm--Klemm--Str.~9, 48149 M\"unster, 
 Germany}
\newcommand*{\ASWarsH}{High Energy Physics Department, National Centre for 
 Nuclear Research, ul.\ Hoza~69, 00-681, Warsaw, Poland}
\newcommand*{\IITB}{Department of Physics, Indian Institute of Technology 
 Bombay, Powai, Mumbai--400076, Maharashtra, India}
\newcommand*{\IKPJ}{Institut f\"ur Kernphysik, Forschungszentrum J\"ulich, 
 52425 J\"ulich, Germany}
\newcommand*{\JCHP}{J\"ulich Center for Hadron Physics, Forschungszentrum 
 J\"ulich, 52425 J\"ulich, Germany}
\newcommand*{\Bochum}{Institut f\"ur Experimentalphysik I, Ruhr--Universit\"at 
 Bochum, Universit\"atsstr.~150, 44780 Bochum, Germany}
\newcommand*{\ZELJ}{Zentralinstitut f\"ur Engineering, Elektronik und 
 Analytik, Forschungszentrum J\"ulich, 52425 J\"ulich, Germany}
\newcommand*{\Erl}{Physikalisches Institut, 
 Friedrich--Alexander--Universit\"at Erlangen--N\"urnberg, 
 Erwin--Rommel-Str.~1, 91058 Erlangen, Germany}
\newcommand*{\ITEP}{Institute for Theoretical and Experimental Physics, State 
 Scientific Center of the Russian Federation, Bolshaya Cheremushkinskaya~25, 
 117218 Moscow, Russia}
\newcommand*{\Giess}{II.\ Physikalisches Institut, 
 Justus--Liebig--Universit\"at Gie{\ss}en, Heinrich--Buff--Ring~16, 
 35392 Giessen, Germany}
\newcommand*{\IITI}{Department of Physics, Indian Institute of Technology 
 Indore, Khandwa Road, Indore--452017, Madhya Pradesh, India}
\newcommand*{\HepGat}{High Energy Physics Division, Petersburg Nuclear Physics 
 Institute, Orlova Rosha~2, Gatchina, Leningrad district 188300, Russia}
\newcommand*{\HISKP}{Helmholtz--Institut f\"ur Strahlen-- und Kernphysik, 
 Rheinische Friedrich--Wilhelms--Universit\"at Bonn, Nu{\ss}allee~14--16, 
 53115 Bonn, Germany}
\newcommand*{\HiJINR}{Veksler and Baldin Laboratory of High Energiy Physics, 
 Joint Institute for Nuclear Physics, Joliot--Curie~6, 141980 Dubna, Moscow 
 region, Russia}
\newcommand*{\Katow}{August Che{\l}kowski Institute of Physics, University of 
 Silesia, Uniwersytecka~4, 40-007, Katowice, Poland}
\newcommand*{\IFJ}{The Henryk Niewodnicza{\'n}ski Institute of Nuclear 
 Physics, Polish Academy of Sciences, 152~Radzikowskiego St, 31-342 
 Krak\'{o}w, Poland}
\newcommand*{\NuJINR}{Dzhelepov Laboratory of Nuclear Problems, Joint 
 Institute for Nuclear Physics, Joliot--Curie~6, 141980 Dubna, Moscow region, 
 Russia}
\newcommand*{\KEK}{High Energy Accelerator Research Organisation KEK, Tsukuba, 
 Ibaraki 305--0801, Japan}
\newcommand*{\IMPCAS}{Institute of Modern Physics, Chinese Academy of 
 Sciences, 509 Nanchang Rd., Lanzhou 730000, China}
\newcommand*{\ASLodz}{Department of Cosmic Ray Physics, National Centre for 
 Nuclear Research, ul.\ Uniwersytecka~5, 90--950 {\L}\'{o}d\'{z}, Poland}

\newcommand*{\Delhi}{Department of Physics and Astrophysics, University of 
 Delhi, Delhi--110007, India}
\newcommand*{\SU}{Department of Physics, Stockholm University, 
 Roslagstullsbacken~21, AlbaNova, 10691 Stockholm, Sweden}
\newcommand*{\Mainz}{Institut f\"ur Kernphysik, Johannes 
 Gutenberg--Universit\"at Mainz, Johann--Joachim--Becher Weg~45, 55128 Mainz, 
 Germany}
\newcommand*{\UCLA}{Department of Physics and Astronomy, University of 
 California, Los Angeles, California--90045, U.S.A.}
\newcommand*{\Bern}{Albert Einstein Center for Fundamental Physics, 
 Fachbereich Physik und Astronomie, Universit\"at Bern, Sidlerstr.~5, 
 3012 Bern, Switzerland}

\author{P.~Adlarson}    \affiliation{\IKPUU}
\author{W.~Augustyniak} \affiliation{\ASWarsN}
\author{W.~Bardan}      \affiliation{\IPJ}
\author{M.~Bashkanov}   \affiliation{\PITue}\affiliation{\Kepler}
\author{F.S.~Bergmann}  \affiliation{\MS}
\author{M.~Ber{\l}owski}\affiliation{\ASWarsH}
\author{H.~Bhatt}       \affiliation{\IITB}
\author{M.~B\"uscher}   \affiliation{\IKPJ}\affiliation{\JCHP}
\author{H.~Cal\'{e}n}   \affiliation{\IKPUU}
\author{I.~Ciepa{\l}}   \affiliation{\IPJ}
% ...

\collaboration{WASA-at-COSY Collaboration}\noaffiliation






\begin{abstract}

Exclusive measurements of the quasi-free $pn \to pp\pi^0\pi^-$ reaction have
been performed by means of $pd$ collisions at $T_p$ = 1.2 GeV using the WASA
detector setup at COSY. Total and differential cross sections have been
obtained covering the energy region $\sqrt s$ = (2.35 - 2.46) GeV, which
includes the region of the ABC effect and its associated resonance
structure. No ABC effect, {\it i.e.} low-mass enhancement is found in the
$\pi^0\pi^-$-invariant mass spectrum -- in agreement with the constraint from
Bose statistics that the isovector pion pair can not be in relative $s$-wave. 
At the upper end of the covered energy region $t$-channel processes for Roper,
$\Delta(1600)$ and $\Delta\Delta$ excitations provide  a reasonable
description of the data, but at low energies the measured cross sections are
much larger than predicted by such processes. Adding a resonance amplitude for
the resonance at $m$~=~2.37 GeV with $\Gamma$ ~=~70 MeV and $I(J^P)~=~0(3^+)$
observed recently in $pn \to d\pi^0\pi^0$ and $pn \to d\pi^+\pi^-$ reactions
leads to an agreement with the data also at low energies.
 
\end{abstract}

\maketitle

\section{Introduction}

Recent data on the basic double-pionic fusion reactions $pn \to d\pi^0\pi^0$
and $pn \to d\pi^+\pi^-$ demonstrate that the so-called ABC effect is tightly
correlated with a narrow resonance structure in the total cross section of
this reaction \cite{prl2011,MB,isofus}. The ABC effect denoting a huge
low-mass enhancement in the $\pi\pi$ invariant mass spectrum is observed to
happen, if the initial 
nucleons or light nuclei fuse to a bound final nuclear system and if the
produced pion pair is isoscalar. Since as of yet no quantitative understanding
of this phenomenon has been available, it has been named after the initials of
Abashian, Booth and Crowe, who first observed it in the inclusive measurement
of the $pd \to ^3$HeX reaction more than fifty years ago \cite{abc}.

The resonance structure with $I(J^P) = 0(3^+)$ \cite{prl2011} observed in the
$pn \to d\pi\pi$ total cross section at $\sqrt s$~=~2.37 GeV is situated about
90 MeV below $\sqrt s = 2 m_{\Delta}$, the peak position of the conventional
$t$-channel $\Delta\Delta$ process, and has a width of only 70 MeV, which is
about three times narrower than this process. From 
 the Dalitz plots of the $pn \to d\pi^0\pi^0$ reaction it is concluded that this
resonance must decay nevertheless via the intermediate $\Delta^+\Delta^0$
system into its final $d\pi^0\pi^0$ state.

If this scenario is correct, then also the $pn \to pp\pi^0\pi^-$ reaction
should be affected by this resonance, 
since this channel may proceed via the same intermediate $\Delta^+\Delta^0$
system. From isospin coupling we expect that the resonance effect in the
$pp\pi^0\pi^-$ system should be half that in the $np\pi^0\pi^0$ system. And
from the estimations in Refs. \cite{col,oset} we expect the resonance effect in
the $np\pi^0\pi^0$ channel to be about 85$\%$ of that in the $d\pi^0\pi^0$ system. Since the
peak resonance cross section in the latter is 270 $\mu$b \cite{isofus}
sitting upon some background due to conventional $t$-channel Roper and
$\Delta\Delta$ excitations, we
estimate the peak resonance contribution in the $pp\pi^0\pi^-$ system to be in
the order of 100 $\mu$b. 

In the following we will demonstrate that in this particular reaction the
resonance is not correlated with the ABC effect for two reasons. First, the
isovector $\pi\pi$ system here is not in relative $s$-wave, but in relative
$p$-wave. And second, in case of unbound nucleons in the final state the form
factor introduced for the description of the ABC-effect in Ref. \cite{prl2011}
does not act on the pions primarily, but on the nucleons. 

Henceforth we will denote the resonance structure by $d^*$ -- following its
notation  
in Refs. \cite{goldman,ping}, where a resonance with the same quantum
numbers has been predicted at just about the mass, where we see this
particular resonance structure. 
Actually, the first prediction of such a resonance dates
back to Dyson and Xuong \cite{Dyson} ($D_{03}$ in their
nomenclature) postulating a mass amazingly close to the one we observe
now. Also, a very recent fully relativistic three-body calculation 
of Gal and Garzilaco \cite{GG} finds this resonance at exactly the position we
observe. For a recent review of the dibaryon issue see Ref. \cite{ST}.

Since in the reaction of interest here the pion pair is produced in the $\rho$
channel, it provides also unique access to the
question, whether this resonance can contribute to $\rho$ production and thus
to $e^+e^-$ production in $np$ collisions. Known as the so-called DLS puzzle
the dilepton production at $T_p \approx$ 1.2 GeV is strongly enhanced in the
mass range 0.3 GeV/c$^2 \leq M_{e^+e^-} \leq$ 0.6 GeV/c$^2$ compared to what is
expected from a conventional reaction scenario, whereas the $pp$
induced dilepton production is in agreement with it \cite{hades}. As a
possible solution of this puzzle $e^+e^-$ production via the $d^*$ resonance
has been proposed \cite{AF}. In fact, first simulations of this resonance
scenario are very promising \cite{dls}, if the $d^*$ production in the
$pp\pi^0\pi^-$ channel turns to be, indeed, in the order of 100 $\mu$b.

Finally, we note that this basic two-pion production reaction has been looked
at so far only by low-statistics bubble-chamber measurements. As a result
there exist no data on  differential observables, just total cross sections at
a few energies \cite{KEK,brunt,dakhno}. Therefore not only from the aspect of
resonance search it appears desirable to collect high-quality data for this
reaction channel, but also from the more general aspect to investigate, to
what extent this reaction channel can be understood by conventional reaction
mechanisms, which have been shown to work well for all $pp$ induced two-pion
production channels -- see discussion section below.

\section{Experiment}

In order to investigate this reaction in more detail experimentally, we have
analyzed a $pd$ 
run at $T_p$ = 1.2 GeV taken in 2009 with the WASA detector facility at COSY
using the deuterium pellet target \cite{barg,wasa}. The hardware trigger
utilized in this analysis requested at least one charged hit in the forward
detector as well as two neutral hits in the central detector. 

The quasi-free reaction $pd \to pp \pi^0\pi^- + p_{spectator}$
has been selected by requiring two proton tracks in the forward detector, an
$\pi^-$ track in the central detector as well as two photons originating from
a $\pi^0$ decay. That way the non-measured proton spectator four-momentum
could be reconstructed by a kinematic fit with two over-constraints. 

In Fig. 1
the reconstructed spectator momentum distribution is shown in comparison with a
Monte-Carlo (MC) simulation of the quasifree $pd \to pp \pi^0\pi^- +
p_{spectator}$ process. The good agreement provides confidence that the data
indeed reflect a quasifree process. As in Ref. \cite{prl2011} we only use
spectator momenta $p_{spectator} <$ 0.16 GeV/c for the further data
analysis. This implies 
an energy range of 2.35 GeV $\leq \sqrt s \leq$ 2.46 GeV being covered due to
the Fermi motion of the nucleons in the target deuteron. This energy range
corresponds to lab incident energies of 1.07 GeV $< T_p < $ 1.36 GeV.

\begin{figure} 
\centering
\includegraphics[width=0.89\columnwidth]{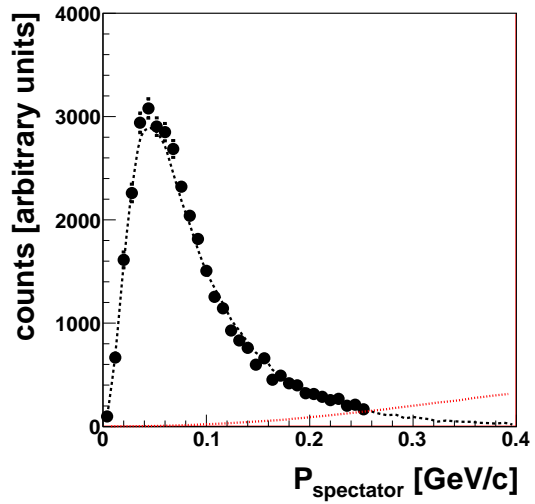}
\caption{\small (Color online)
  Distribution of the spectator proton momenta in the $pd \to pp\pi^0\pi^- + 
  p_{spectator}$ reaction. Data are given by solid dots. The dashed line shows
  the expected distribution for the quasifree process based on the CD Bonn 
  potential \cite{mach} deuteron wavefunction. For comparison the dotted line
  gives the pure phase-space distribution as expected for a coherent reaction
  process: it extends up to momenta of 1.5 GeV/c and peaks around 0.7
  GeV/c. For the data analysis only events with $p_{spectator} <$ 0.16 GeV/c
  have been used.
}
\label{fig1}
\end{figure}

In total a sample of about 42000 good events has been selected.
The requirement that the two protons have to be in the angular range covered
by the forward detector and that the $\pi^-$ and the gammas resulting from
$\pi^0$ decay have to be in the angular range of the central detector reduces
the overall acceptance to about 25$\%$. 
Efficiency and acceptance corrections of the data have been performed by MC
simulations of reaction process and detector setup. For the MC simulations
model descriptions have been used, which will be discussed below in the next
chapter. Since the acceptance is substantially below 100$\%$, the efficiency
corrections are not fully model independent. The error bars in Fig.~2 and the
hatched grey histograms in 
Figs. 3 - 9 give an estimate for systematic uncertainties due to the use of
different models with and without $d^*$ resonance hypothesis for the
efficiency correction.  

The absolute normalization
of the data has been achieved via the simultaneous measurement of the
quasi-free single pion production process $pd \to pp \pi^0 + n_{spectator}$
and comparison of its result to previous bubble-chamber results for the $pp
\to pp \pi^0$ reaction \cite{shim,eis}. That way the uncertainty in the
absolute normalization of our data is that of the previous $pp \to pp \pi^0$
data, {\it i.e.} in the order of 20$\%$.

\section{Results and Discussion}

In order to determine the energy dependence of the total cross section we have
divided our data sample into 10 MeV bins in $\sqrt s$. The resulting total
cross sections together with their statistical and systematic uncertainties
are listed in Table 1.

\begin{table}
\caption{Total cross sections obtained in this work for the $pn \to
  pp\pi^0\pi^-$ reaction in dependence of the center-of-mass energy $\sqrt s$
  and the proton beam energy $T_p$. Systematic uncertainties are given as
  obtained from MC simulations for the detector performance assuming various
  models for the reaction process. } 
\begin{tabular}{llllll} 
\hline

 & $\sqrt s$ & $T_p$ &~~~~$\sigma_{tot}$  &~~~~$\Delta\sigma_{stat}$ &~~~~$\Delta\sigma_{sys}$  \\ 
& [MeV] & [MeV] &~~~~[$\mu$b] &~~~~[$\mu$b]  &~~~~[$\mu$b] \\

\hline

& 2.35 & 1.075 &~~~~~93 &~~~~2 &~~~11 \\
& 2.36 & 1.100 &~~~~124 &~~~~3 &~~~20  \\
& 2.37 & 1.125 &~~~~165 &~~~~3 &~~~29  \\
& 2.38 & 1.150 &~~~~177 &~~~~3 &~~~23  \\ 
& 2.39 & 1.186 &~~~~186 &~~~~3 &~~~21  \\ 
& 2.40 & 1.201 &~~~~195 &~~~~3 &~~~15  \\
& 2.41 & 1.227 &~~~~215 &~~~~3 &~~~17 \\
& 2.42 & 1.253 &~~~~238 &~~~~3 &~~~18 \\
& 2.43 & 1.279 &~~~~278 &~~~~4 &~~~21 \\
& 2.44 & 1.305 &~~~~277 &~~~~5 &~~~21 \\
& 2.45 & 1.331 &~~~~320 &~~~~6 &~~~25 \\
& 2.46 & 1.357 &~~~~397 &~~~~9 &~~~31 \\ 
\hline
 \end{tabular}\\
\end{table}

Fig.~2 exhibits the energy dependence of the total cross section. The result
of this work is given by the full circles and compared to previous
bubble-chamber measurements from KEK (open circles) \cite{KEK}, NIMROD at RAL
(open triangles) \cite{brunt} and Gatchina (open squares) \cite{dakhno}. The
latter are known to give much too high cross sections, see {\it e.g.} the
$pp\pi^+\pi^-$ channel \cite{iso}. Hence we will disregard them for the
further discussion. 
In the overlap region our data agree well with the bubble-chamber results from
KEK and RAL. The data exhibit a smooth energy dependence of a monotonically
rising cross section with no particular evidence for a narrow resonance
structure in the region of the ABC effect around $T_p$ = 1.13 GeV. However, at
closer inspection the data indicate some kind of plateau in just this region. 

\begin{figure} 
\begin{center}
\includegraphics[width=0.5\textwidth]{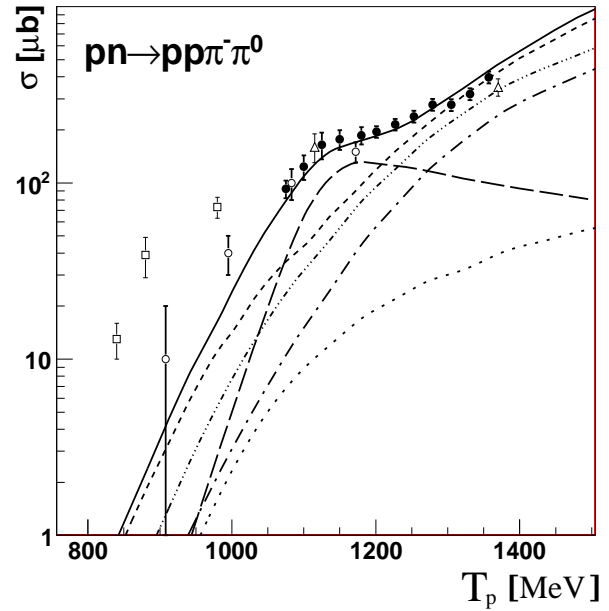}
\caption{ 
   Total cross sections for the $pn \to pp\pi^0\pi^-$ 
  reaction. The results of this work are shown by the full circles together
  with their error bars, which include both statistical and systematic
  uncertainties as given by Table 1. Previous bubble-chamber measurements from
  KEK \cite{KEK} are displayed by open circles, those from NIMROD at RAL
  \cite{brunt} by open triangles and those from Gatchina \cite{dakhno} by open
  squares. The original Valencia model calculations are shown by the
  dot-dot-dot-dashed  curve. Contributions from Roper excitation and its decay
  into $N^* \to \Delta\pi$ are given by the dotted line and those from the
  $t$-channel $\Delta\Delta$ process  by the dash-dotted line. The modified
  Valencia model calculation is shown by the short-dashed line. The solid
  curve shows the result, if the $s$-channel $d^*$ resonance amplitude is
  added. The $d^*$ contribution itself is given by the long-dashed curve.
}
\label{fig2}
\end{center}
\end{figure}

The data are first compared to theoretical calculations in the framework of the
Valencia model \cite{luis}, which incorporates non-resonant and resonant
$t$-channel processes for two-pion production in $NN$ collisions. Resonance
processes concern here the excitation and decay of the $\Delta\Delta$ system
as well as the excitation of the Roper resonance and its subsequent decay
either directly into the $N\pi\pi$ system or via the $\Delta\pi$
system. Compared to the original 
Valencia calculations \cite{luis} the present calculations have been tuned to
describe quantitatively the isovector two-pion production reactions $pp \to
NN\pi\pi$ \cite{iso}, in particular the $pp\pi^0\pi^0$ \cite{deldel} and
$nn\pi^+\pi^+$ \cite{nnpipi} channels by the following modifications:

\begin{itemize}
\item relativistic corrections for the $\Delta$ propagator as given by
  Ref. \cite{ris},
\item strongly reduced $\rho$-exchange contribution in the $t$-channel
  $\Delta\Delta$ process -- in agreement with calculations from Ref. \cite{xu},
\item reduction of the $N^* \to \Delta\pi$ amplitude by a factor of two in
  accordance with $pp \to pp\pi^0\pi^0$ and $pp \to pp\pi^+\pi^-$ measurements
  close to threshold \cite{WB,JP,ae,Roper} as well as in agreement with the
  analysis of photon- and pion-induced pion production on the nucleon
  \cite{boga}, 
\item inclusion of the $t$-channel excitation of the $\Delta(1600)P_{33}$
  resonance.
\end{itemize}

The latter modification was necessary, in order to account for the
unexpectedly large $pp \to nn\pi^+\pi^+$ cross section \cite{nnpipi}. The
predictive power of these modifications has been demonstrated by its
successful application to the recent $pp \to pp\pi^0\pi^0$ data obtained with
WASA at COSY at $T_p$ = 1.4 GeV \cite{TT}. 

Though these
modifications significantly affect the differential distributions, their
effect on the total cross section of the $pn \to pp\pi^0\pi^-$ reaction is
predominantly just in absolute scale -- compare the dot-dot-dot-dashed line in
Fig. 2 with the short-dashed 
one. The dot-dashed line in Fig. 2 denotes the $t$-channel $\Delta\Delta$
process and the dotted line the $t$-channel Roper excitation with subsequent
$N^* \to \Delta\pi$ decay. 

We note by passing that in the energy region of interest the $pp$ final
state interaction is not of importance -- see, {\it e.g.} the
$M_{pp}$ spectrum in Fig.~6, top left of Ref. \cite{deldel}, where the solid
line shown there exhibits only a tiny enhancement at threshold due to the $pp$
final state interaction.  

The original Valencia calculations give cross sections, which are
substantially below the data at all energies. The modified calculations
provide a reasonable description of the data at high energies -- mainly due to
the inclusion of the $\Delta(1600)$ excitation --, but also fail largely at
energies below 1.3 GeV,
where they predict cross sections, which are by as much as a factor of four too
small. Since such a large failure has not been observed in $pp$ induced, {\it
  i.e.} isovector two-pion channels -- and since there is no $t$-channel
resonance process known, which could feed this low-energy region --, the reason
for this striking failure must be in a low-energy two-pion production
process, which is not taken into account in the Valencia model and which has
not much influence on the well-measured $pp$ initiated two-pion production
channels.  

In Ref. \cite{xu} it has been shown that the so-called nucleon-pole term could
possibly be such a process. According to their calculations it provides even the
largest contribution close to threshold in the $pn \to pp\pi^0\pi^-$
reaction. Still, its contribution is far too low to account for these
discrepancies here.  


We conclude that this failure points to an important isoscalar reaction
component, which is not included in the $t$-channel treatment of two-pion
production. It is intriguing that this failure appears to be largest in the
energy region, where the ABC-effect and its associated resonance in the total
cross section have been observed in the isoscalar part of the double-pionic
fusion to deuterium. Hence we add tentatively the amplitude of this resonance
at M = 2.37 GeV and $\Gamma$ = 70 MeV to the conventional amplitude. According
to the consideration in the introduction we have chosen a peak cross section
of 100 $\mu$b for this resonance contribution. It is amazing, how well the
resulting curve (solid line in Fig.2) describes the data. Adjusting the
resonance contribution to the data requires a peak cross section in the
range of 90 - 130 $\mu$b -- depending on the systematic
uncertainties associated with our values for the total cross section.

For a four-body final state there are seven independent differential
observables. We choose to show in this paper the differential distributions
for the invariant masses $M_{\pi^0\pi^-}$, $M_{p\pi^-}$, $M_{pp}$,
$M_{pp\pi^0}$ as well as the differential distributions for the
center-of-mass (cm) angles for protons and pions, namely $\Theta_p^{cm}$,
$\Theta_{\pi^0}^{cm}$ and $\Theta_{\pi^-}^{cm}$. These
distributions are shown in Figs. 3 - 9 with each of them plotted for four
energy bins: 2.35 GeV $ < \sqrt s <$ 2.36 GeV (a), 2.365 $< \sqrt s <$ 2.375
GeV (b), 2.40 $< \sqrt s <$ 2.41 GeV (c) and 2.44 GeV $< \sqrt s <$ 2.45 GeV
(d). 
The second region is chosen to cover just the peak region of the $d^*$ resonance
structure observed in the $pn \to d\pi^0\pi^0$ reaction.

\begin{figure}[t]
\begin{center}

\includegraphics[width=0.23\textwidth]{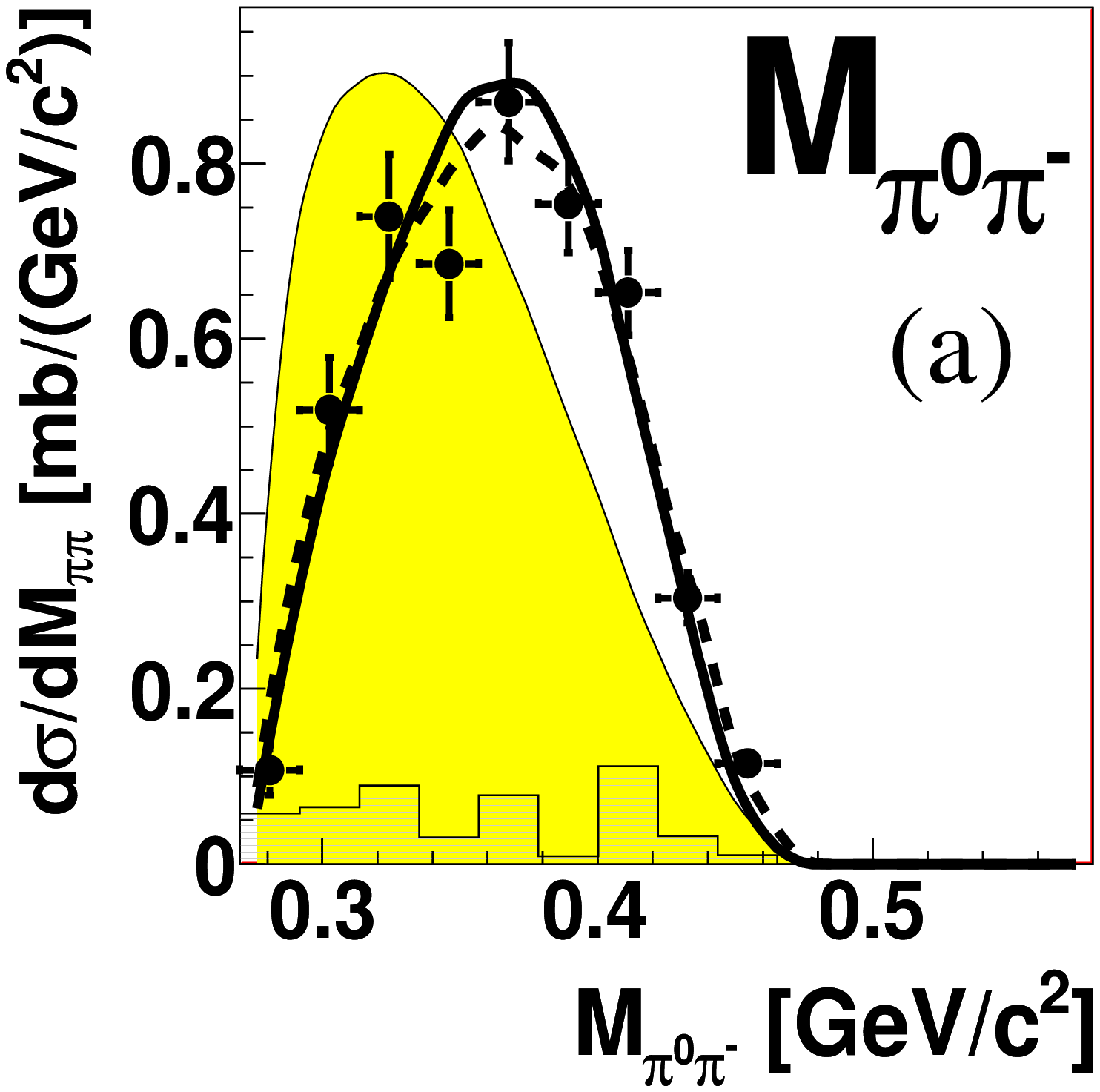}
\includegraphics[width=0.23\textwidth]{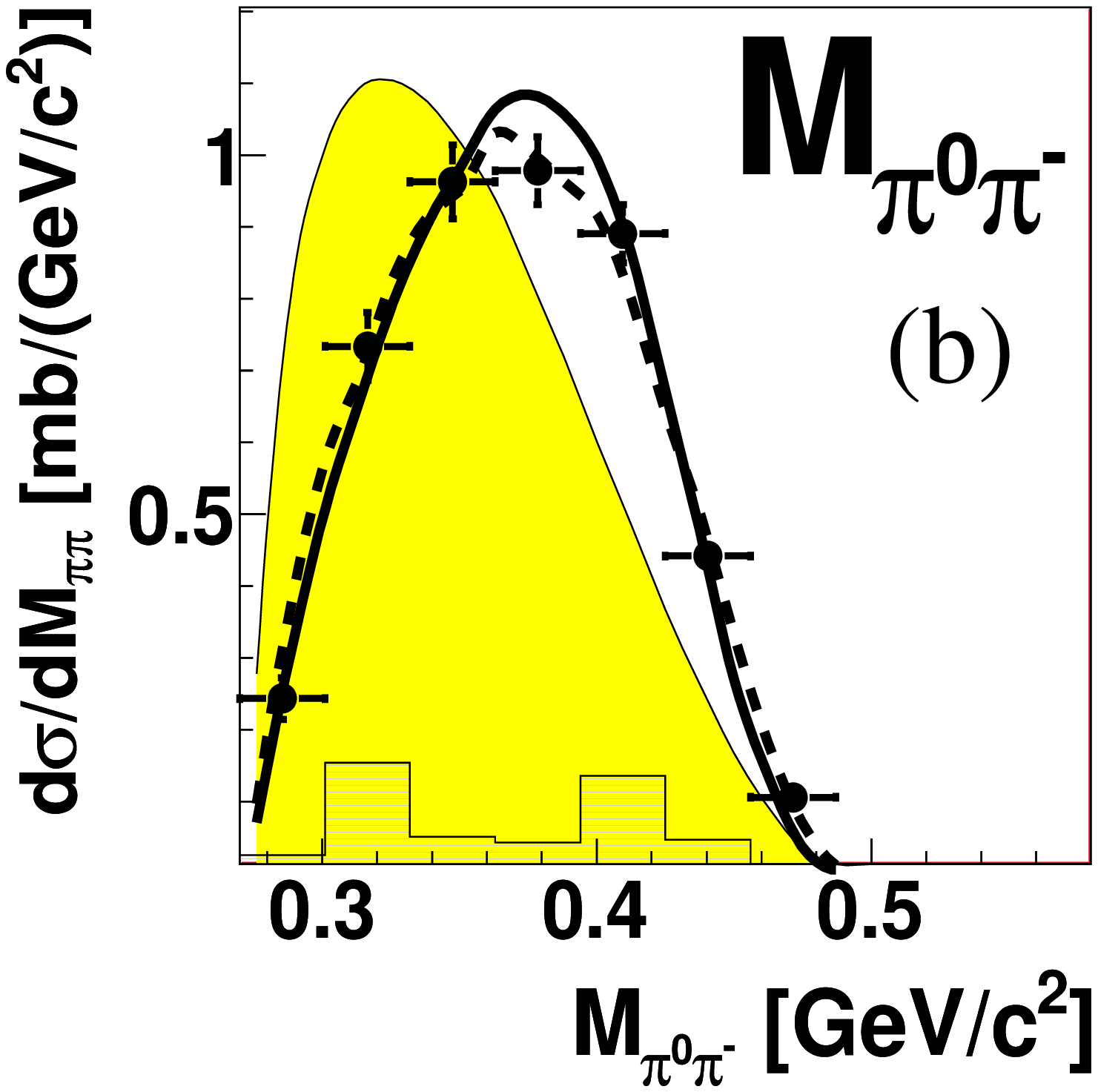}
\includegraphics[width=0.23\textwidth]{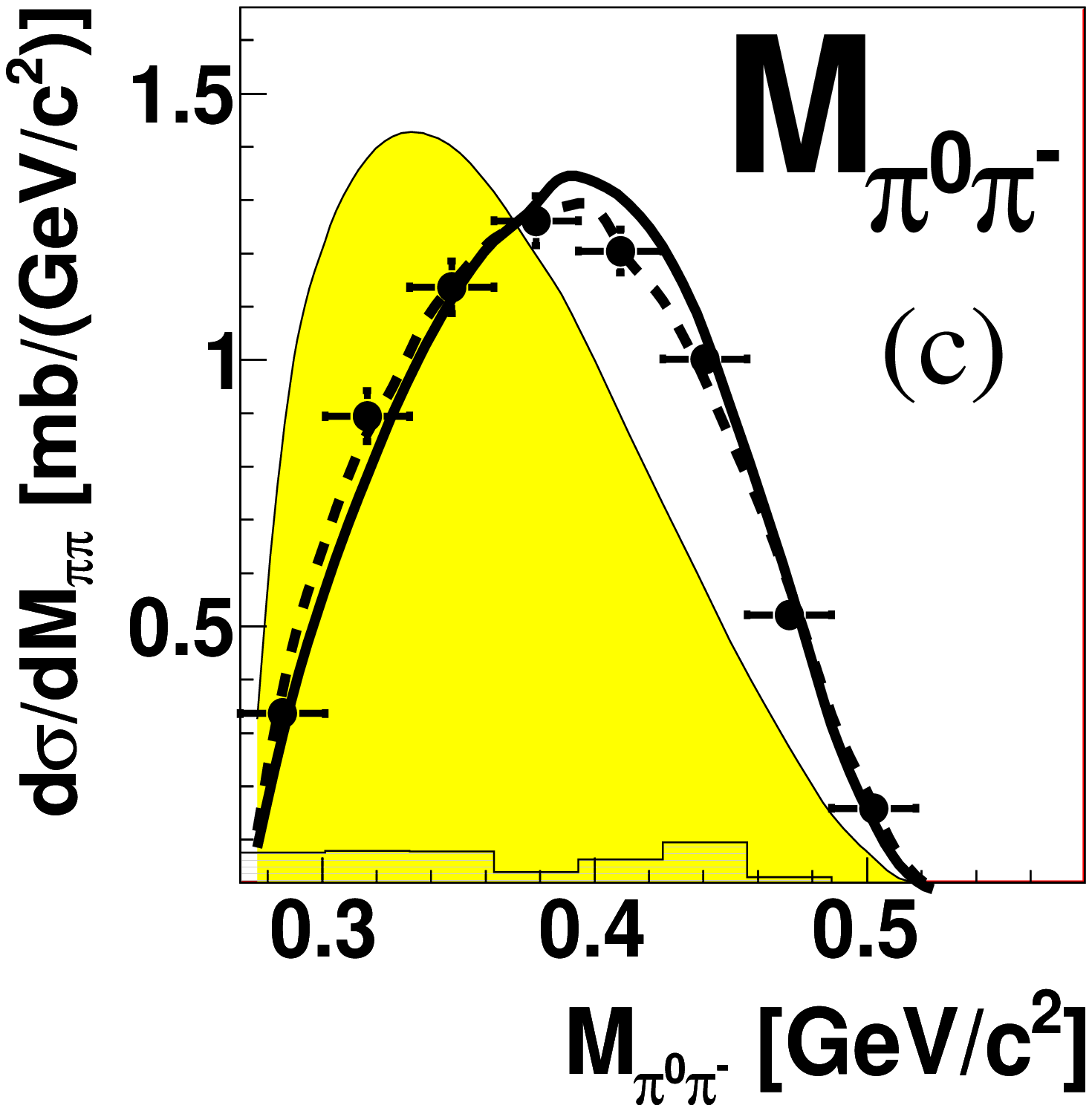}
\includegraphics[width=0.23\textwidth]{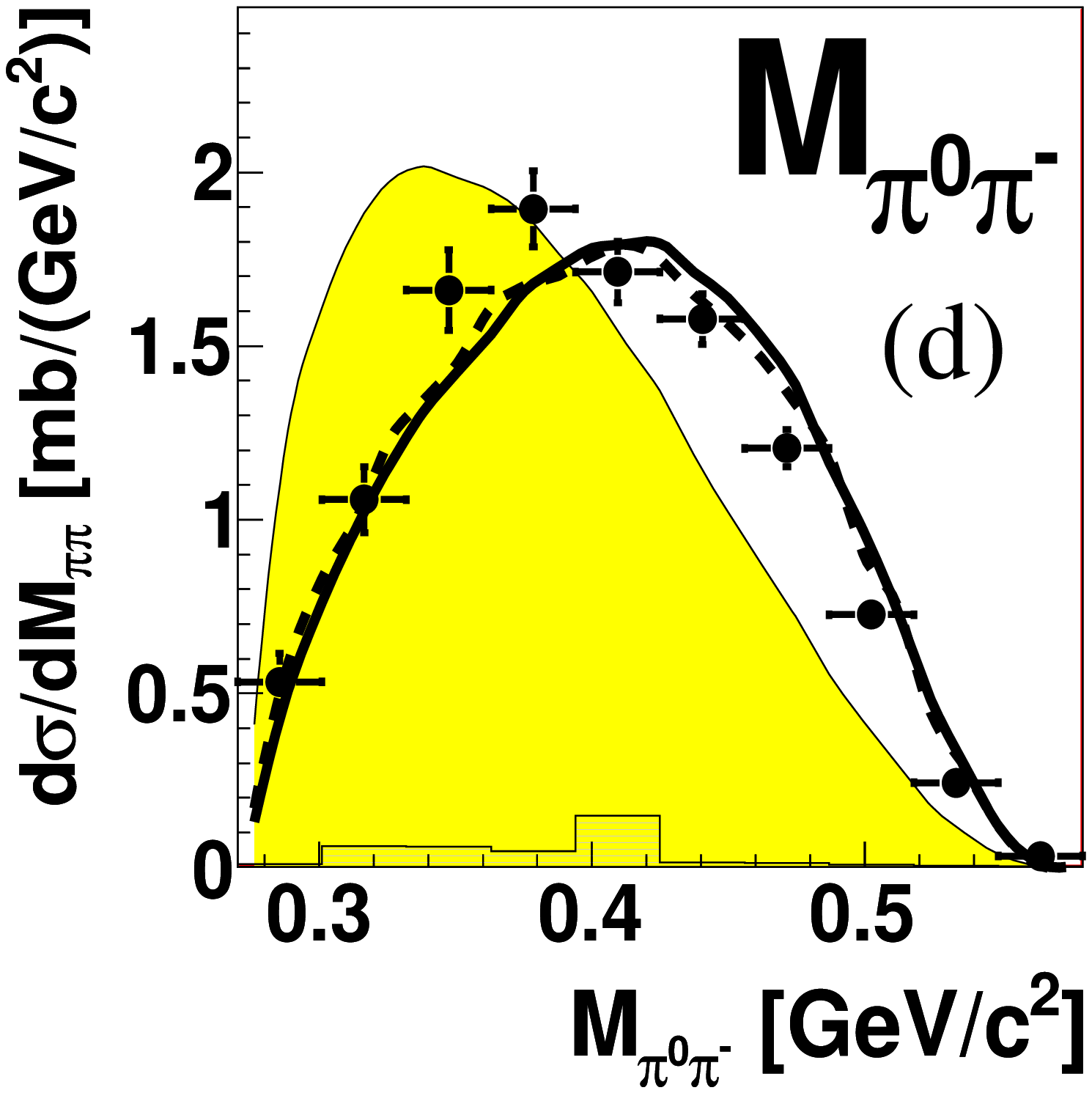}

\caption{(Color online)  
   Distribution of the $\pi^0\pi^-$ invariant mass $M_{\pi^0\pi^-}$ for the $pn
   \to pp\pi^0\pi^-$ reaction at 2.35 GeV $ < \sqrt s <$ 2.36 GeV (a), 2.365 $<
   \sqrt s <$ 2.375 GeV (b), 2.40 $< \sqrt s <$ 2.41 GeV (c) and 2.44 GeV $<
   \sqrt s <$ 2.45 GeV (d) corresponding to beam energy bins 1.07 GeV $< T_p
   <$ 1.10  GeV, 1.11 GeV $< T_p <$ 1.14 GeV, 1.20 GeV $< T_p <$ 1.23 GeV and
   1.30 GeV $< T_p <$ 1.33 GeV. 
   Filled
   circles represent the experimental results of this work. The hatched
   histograms give estimated systematic uncertainties due to the incomplete
   coverage of the solid angle. The shaded
   areas denote phase space distributions. The dashed lines are calculations 
   with the modified Valencia model. The solid lines shows the result, if the
   $d^*$ resonance amplitude is added. 
   All calculations are normalized in area to the data. 
}
\label{fig3}
\end{center}
\end{figure}




\begin{figure}
\begin{center}

\includegraphics[width=0.23\textwidth]{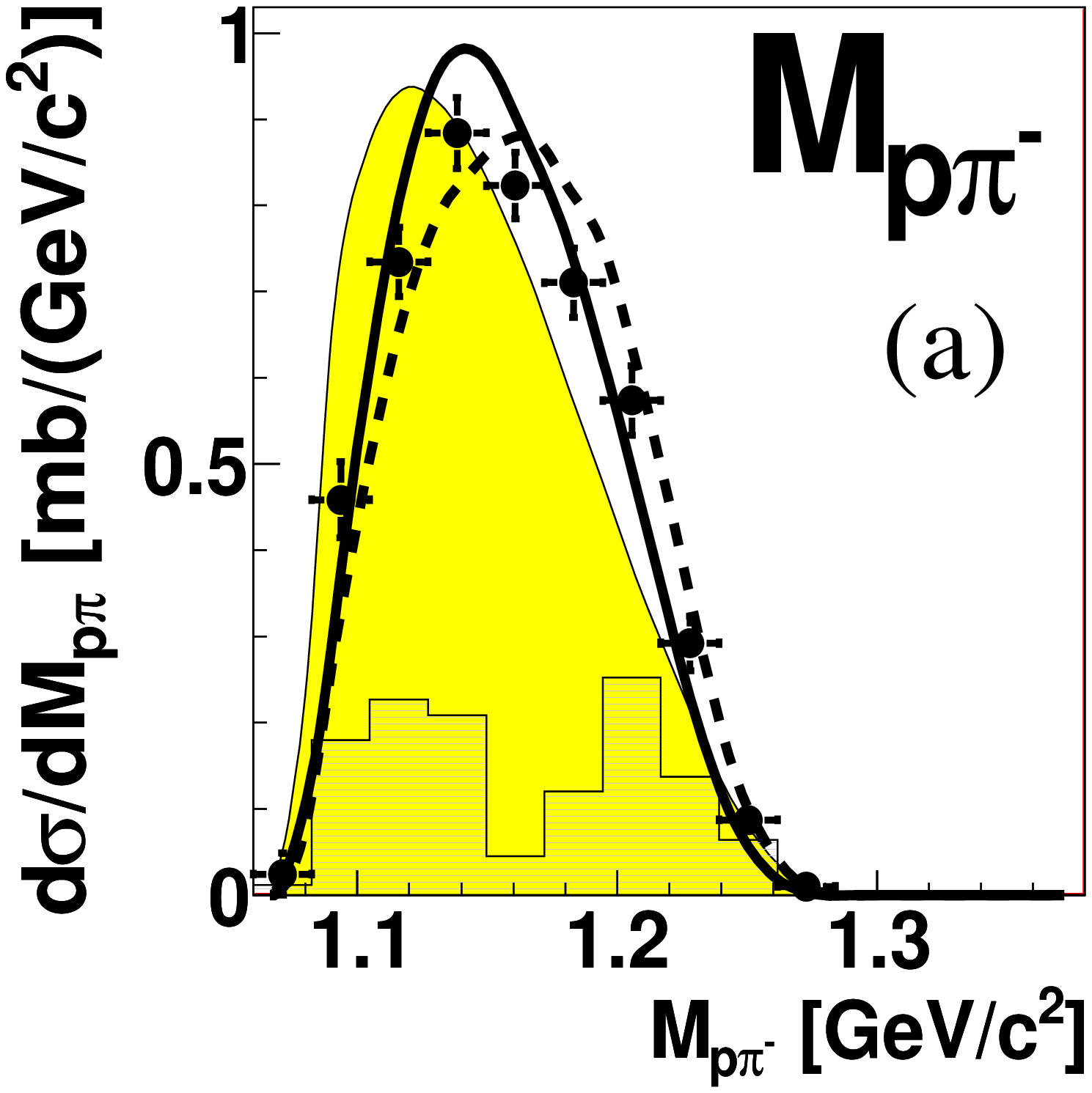}
\includegraphics[width=0.23\textwidth]{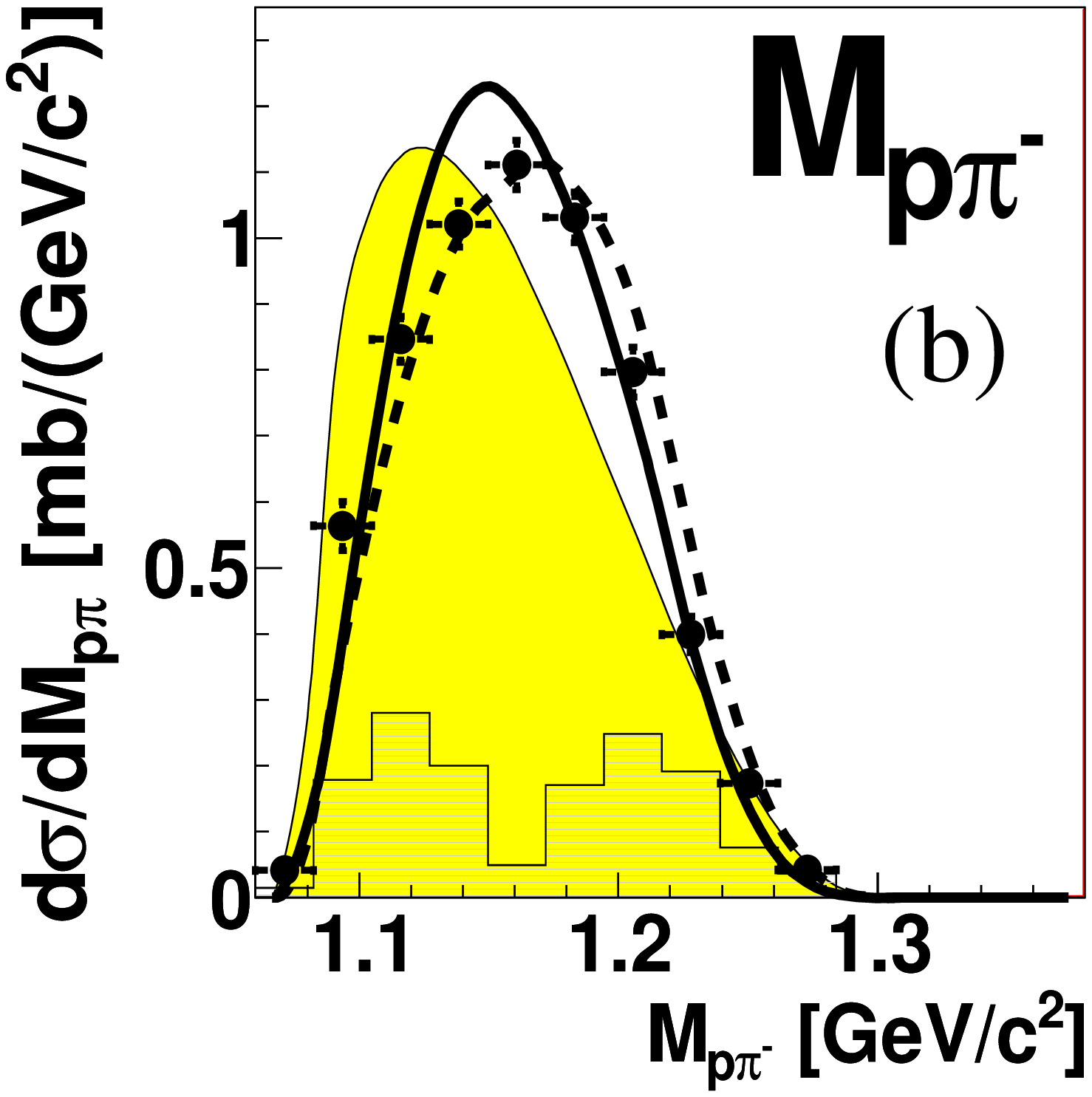}
\includegraphics[width=0.23\textwidth]{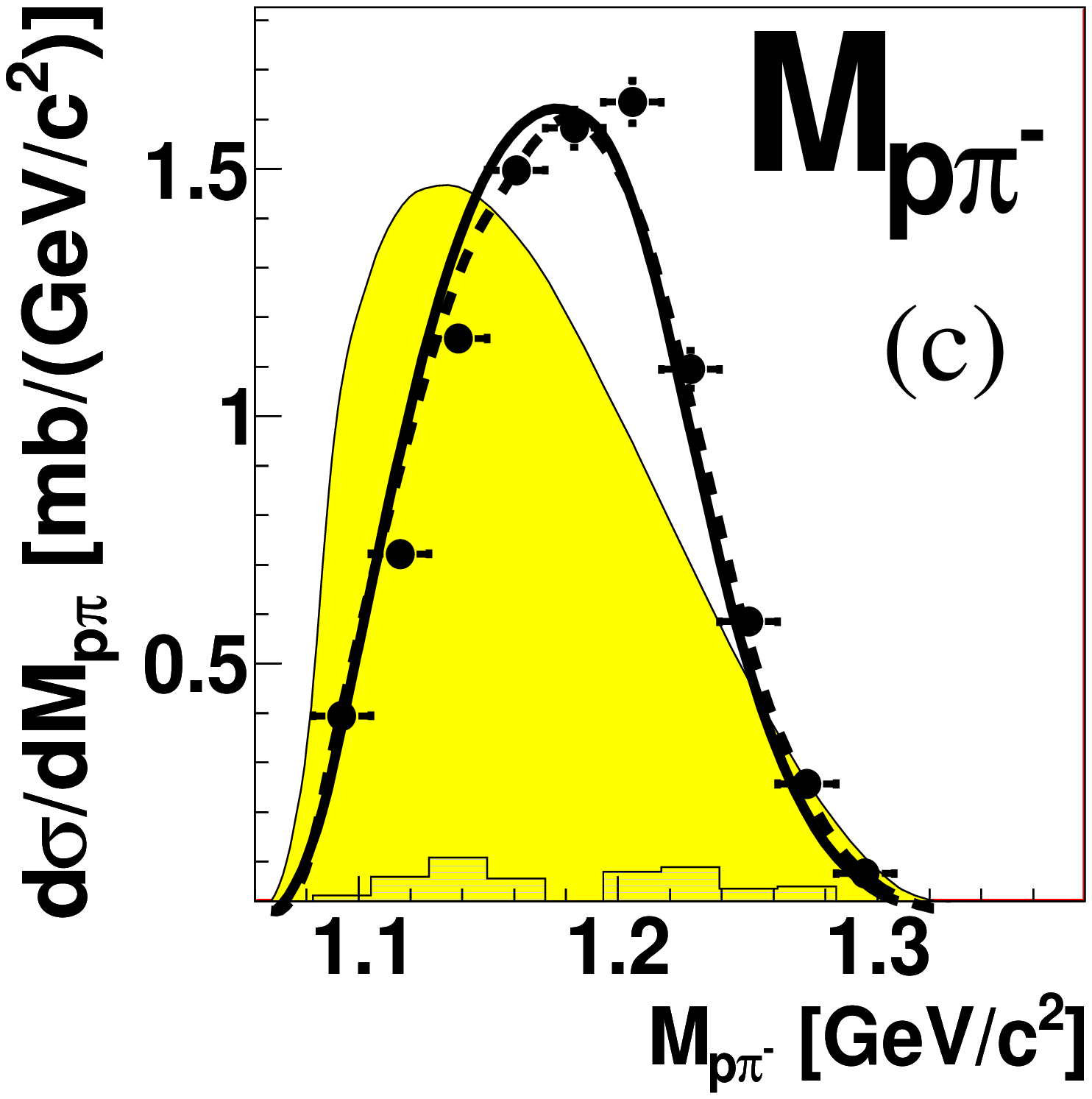}
\includegraphics[width=0.23\textwidth]{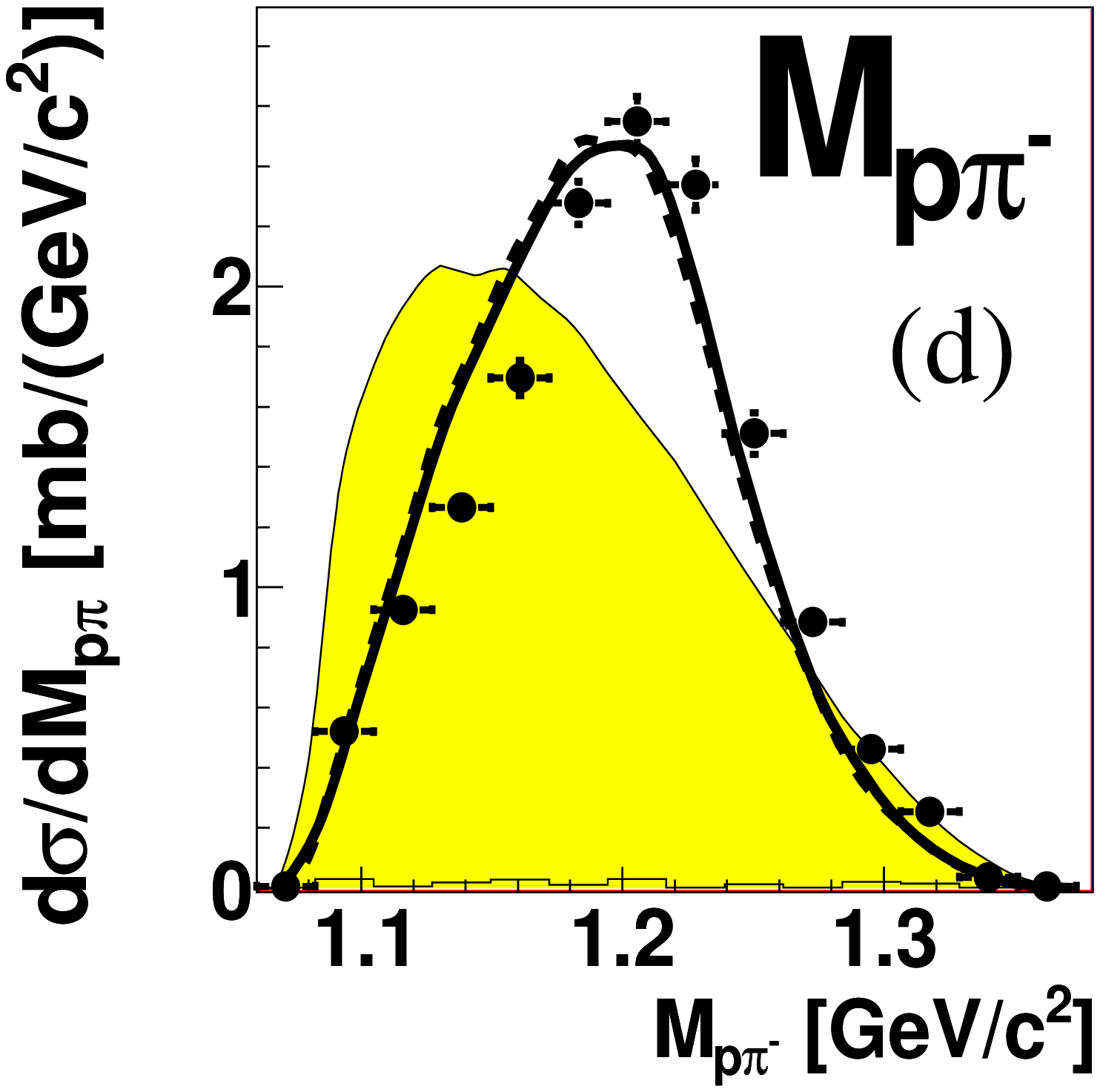}

\caption{(Color online) 
  Same as Fig. 3 but for the 
   distributions of the invariant masses $M_{p\pi^-}$. 
}
\label{fig5}
\end{center}
\end{figure}

\begin{figure}
\begin{center}
\includegraphics[width=0.23\textwidth]{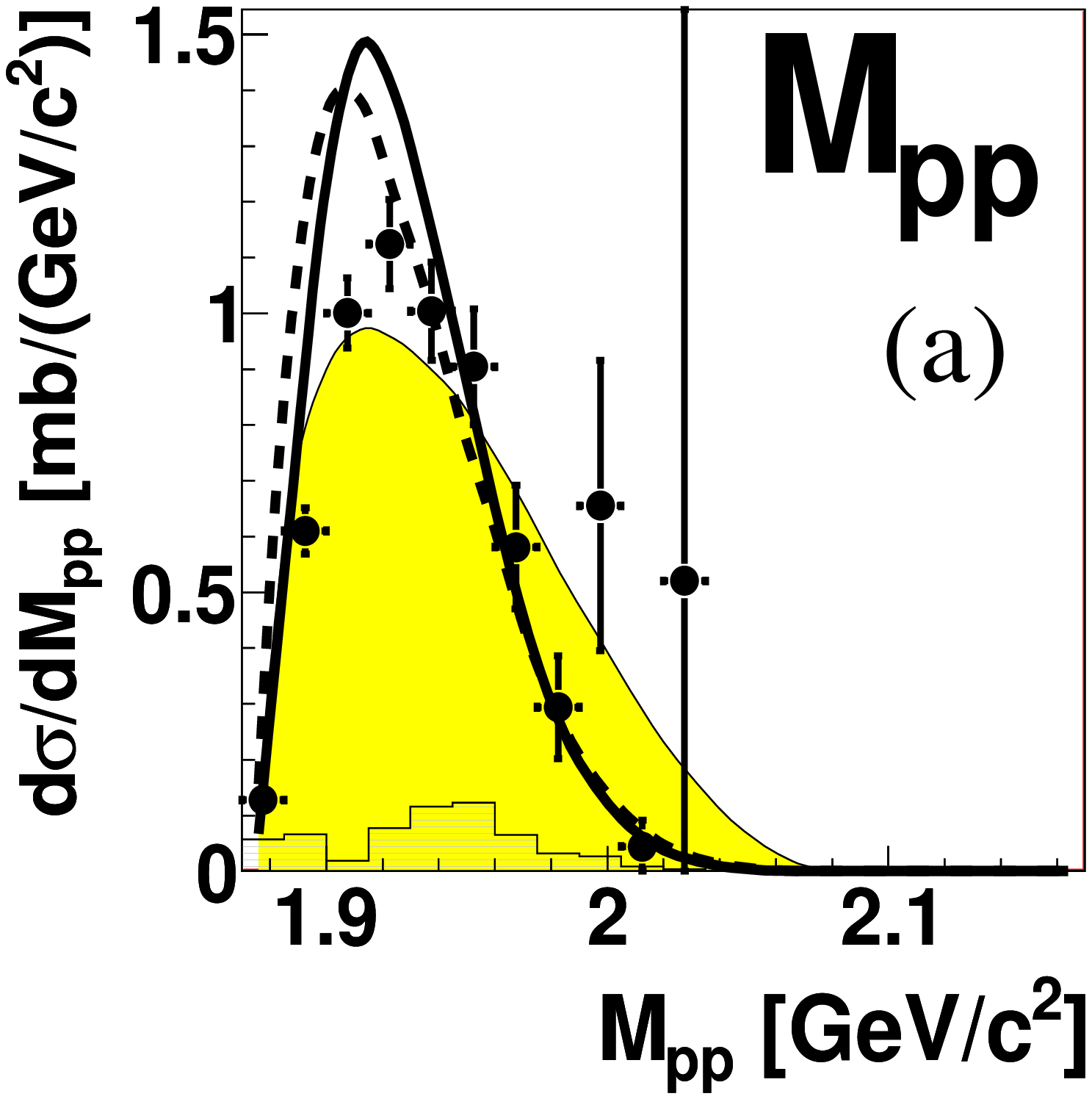}
\includegraphics[width=0.23\textwidth]{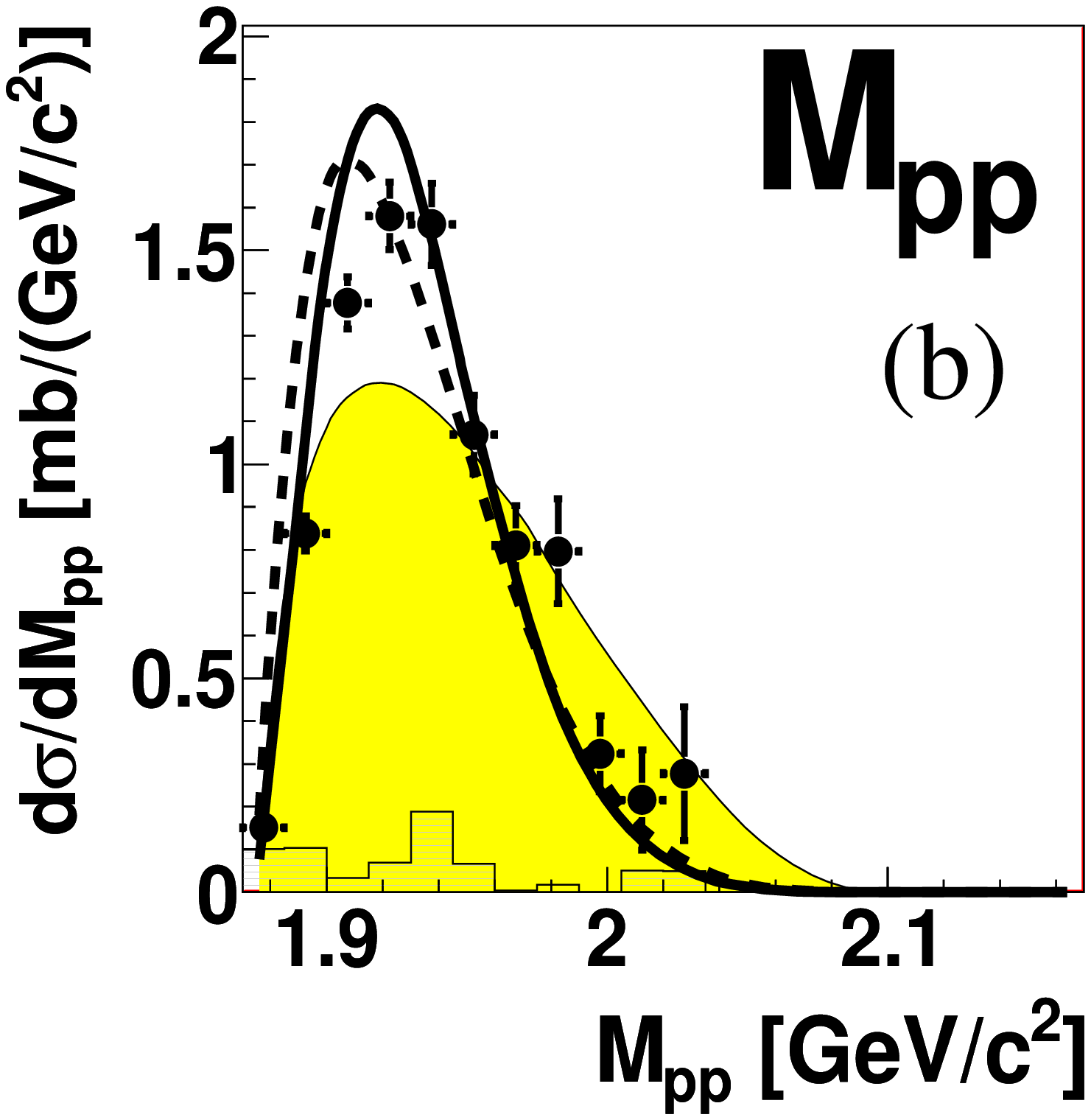}
\includegraphics[width=0.23\textwidth]{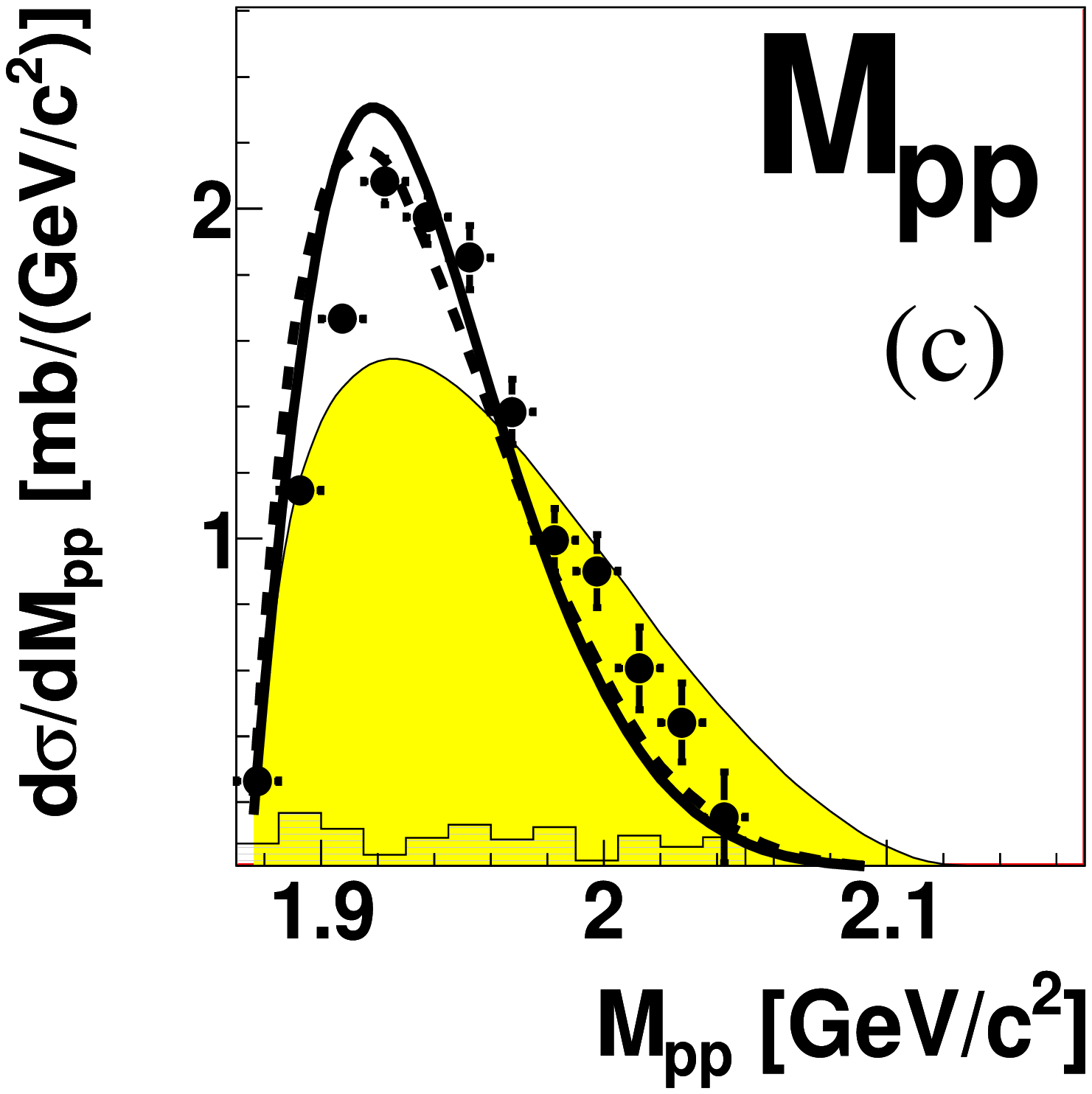}
\includegraphics[width=0.23\textwidth]{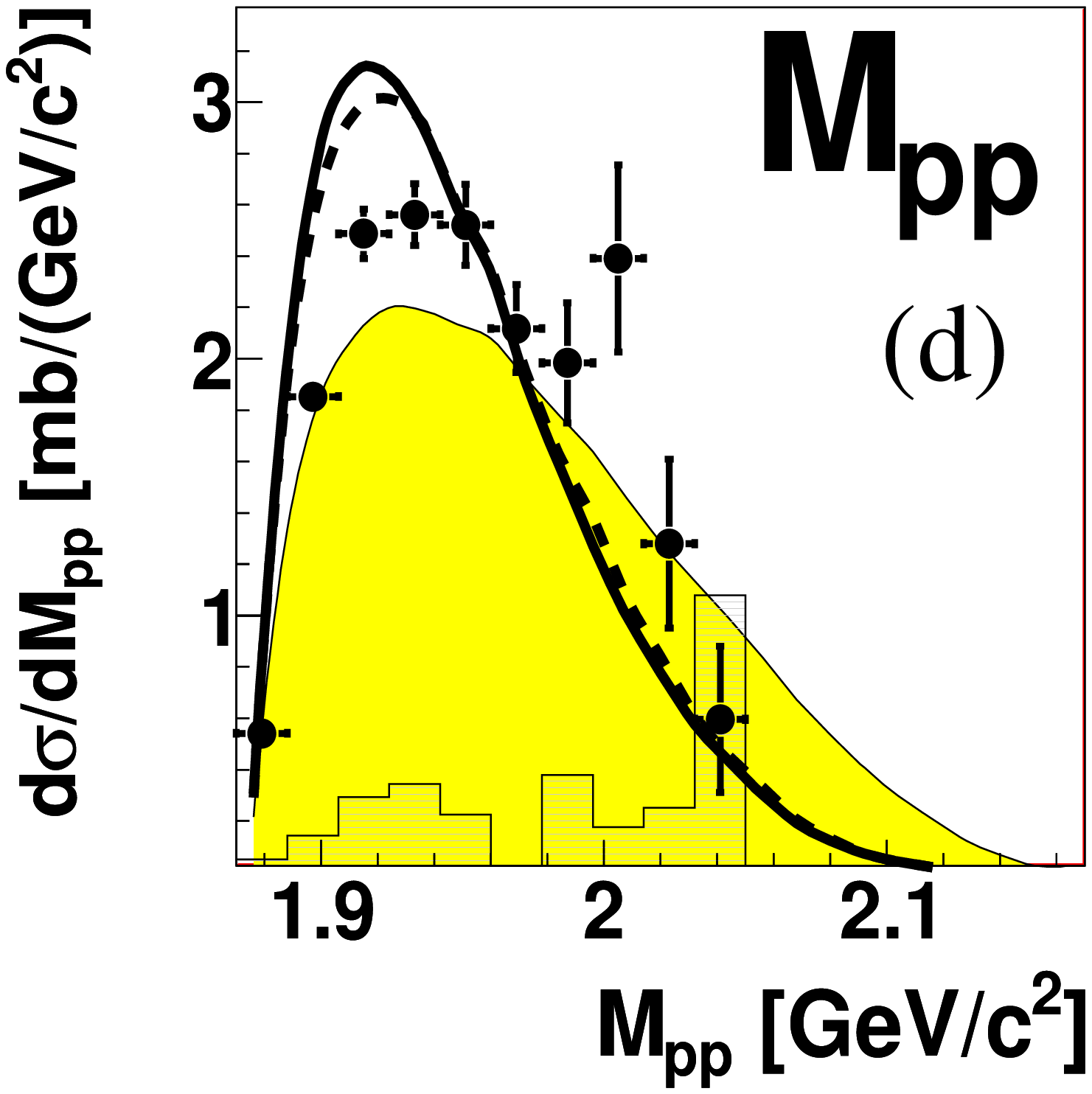}

\caption{(Color online) 
  Same as Fig. 3 but for the 
   distributions of the invariant masses $M_{pp}$. 
}
\label{fig6}
\end{center}
\end{figure}

\begin{figure}
\begin{center}
\includegraphics[width=0.23\textwidth]{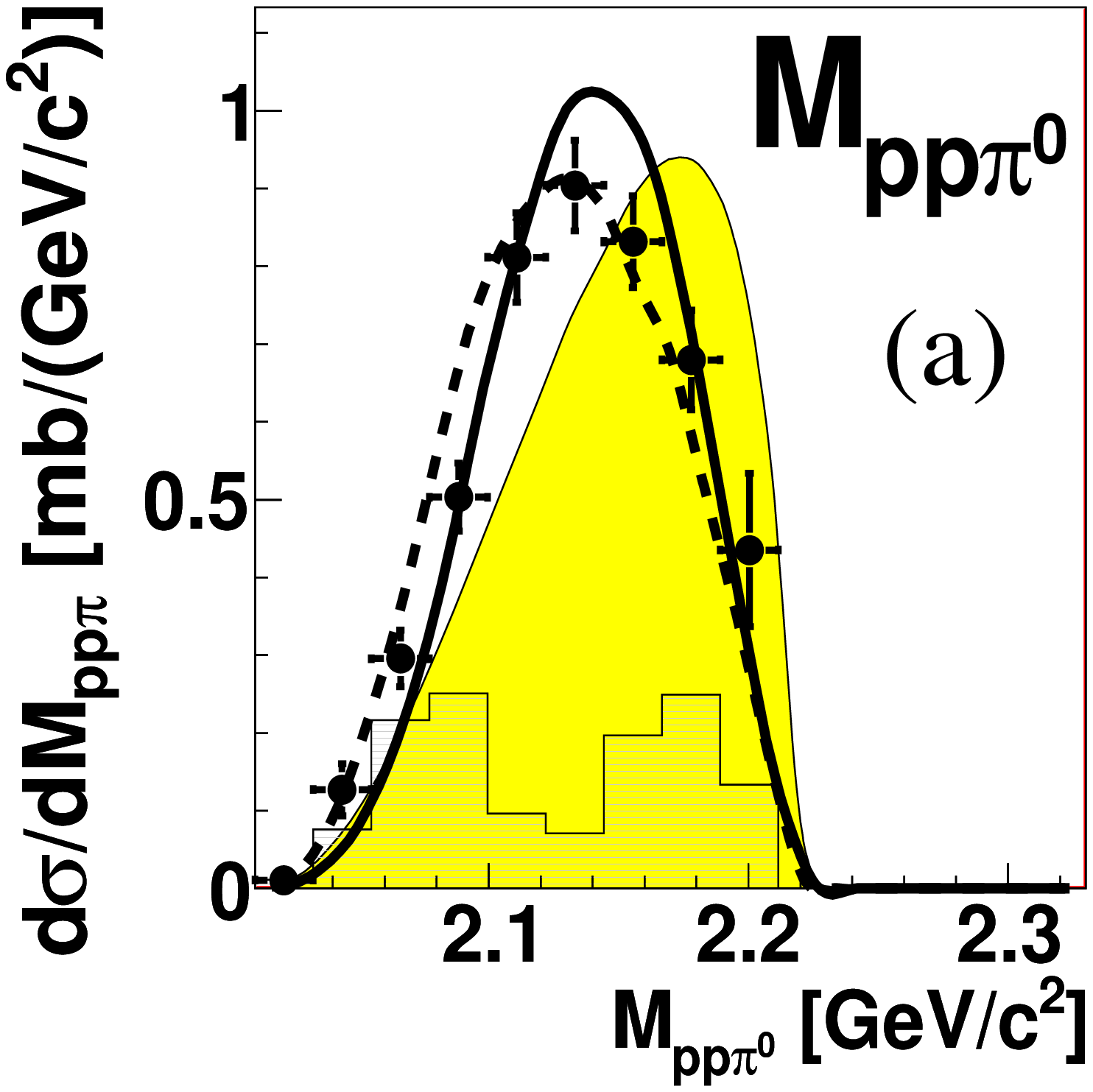}
\includegraphics[width=0.23\textwidth]{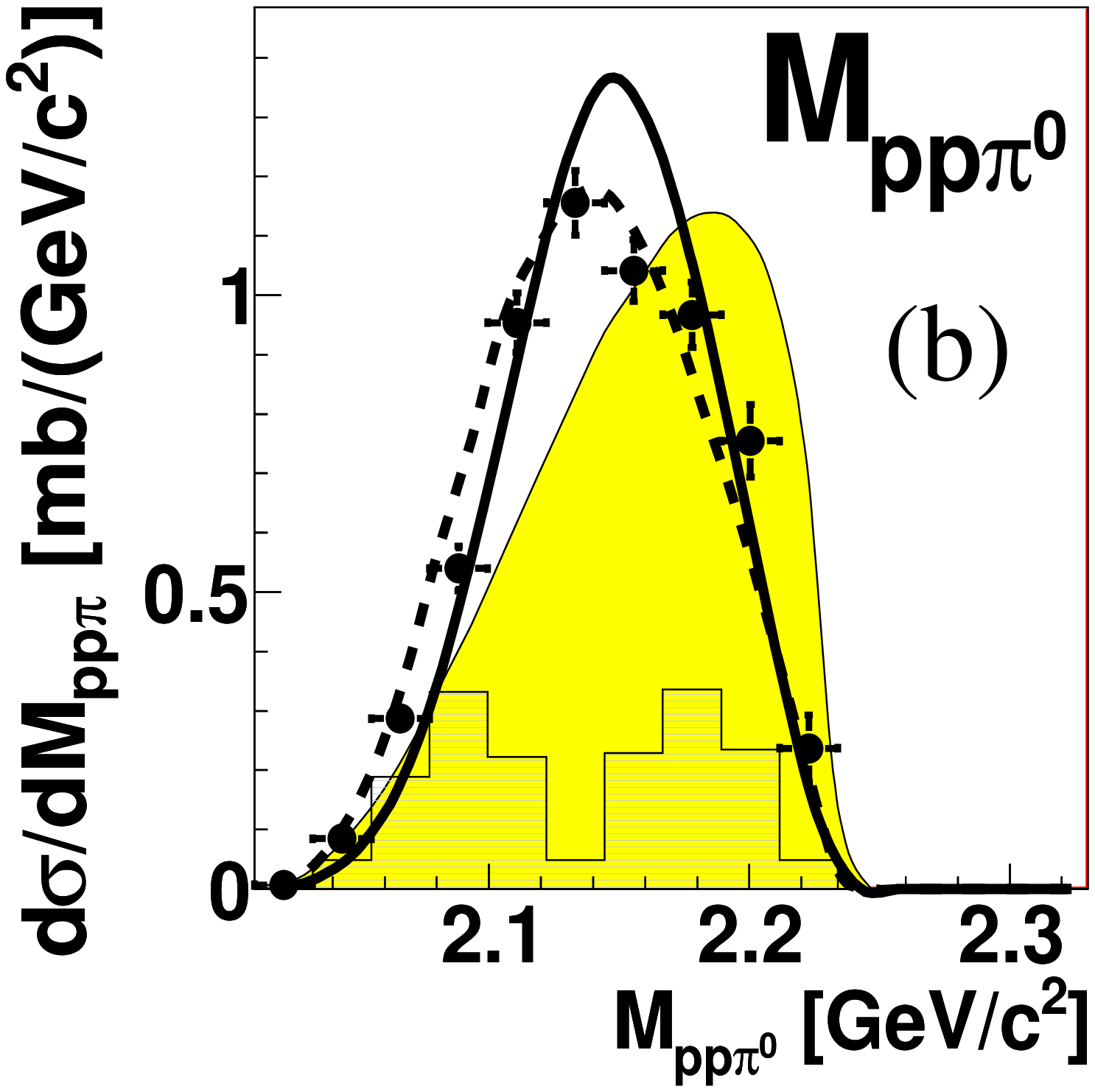}
\includegraphics[width=0.23\textwidth]{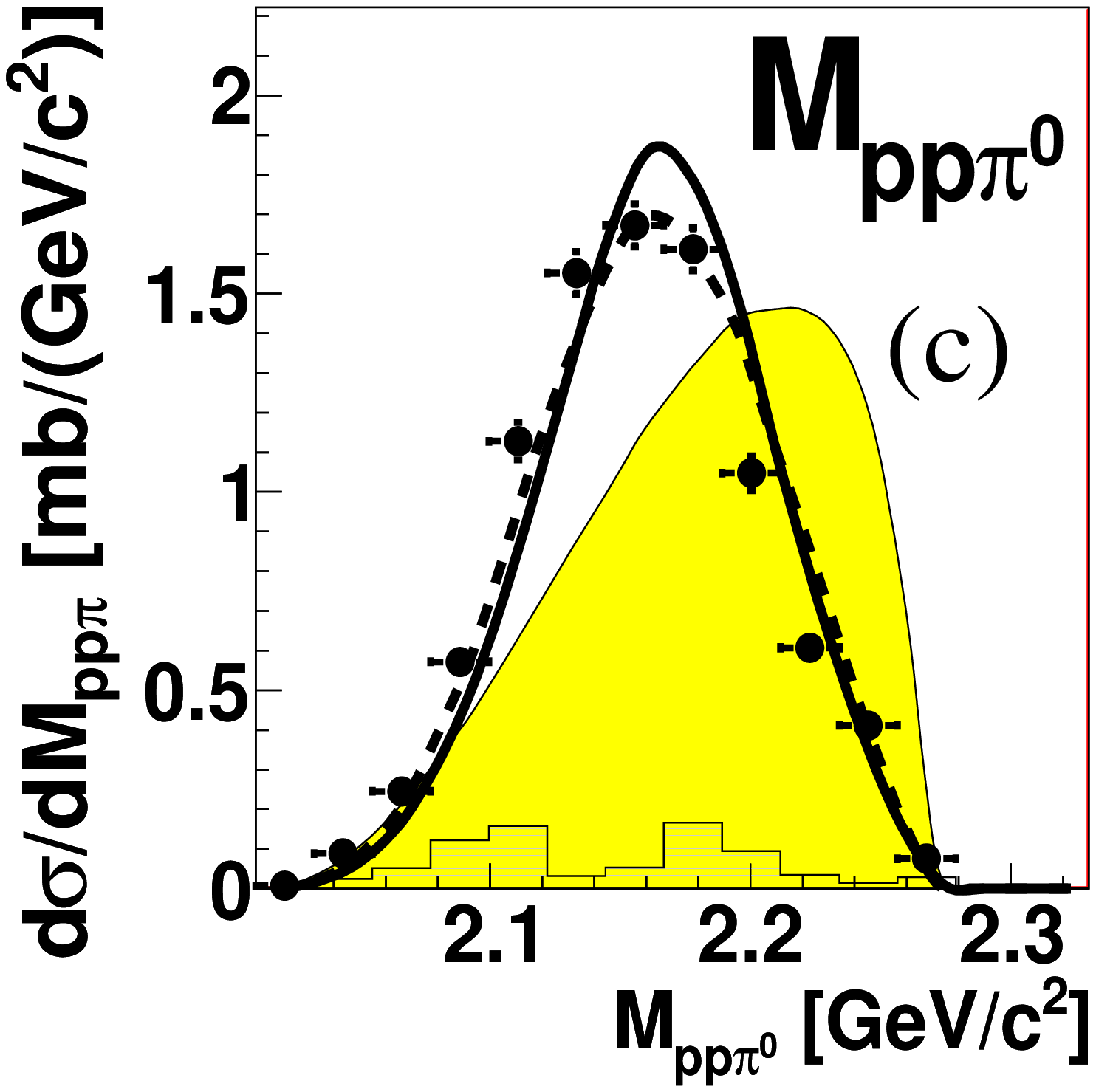}
\includegraphics[width=0.23\textwidth]{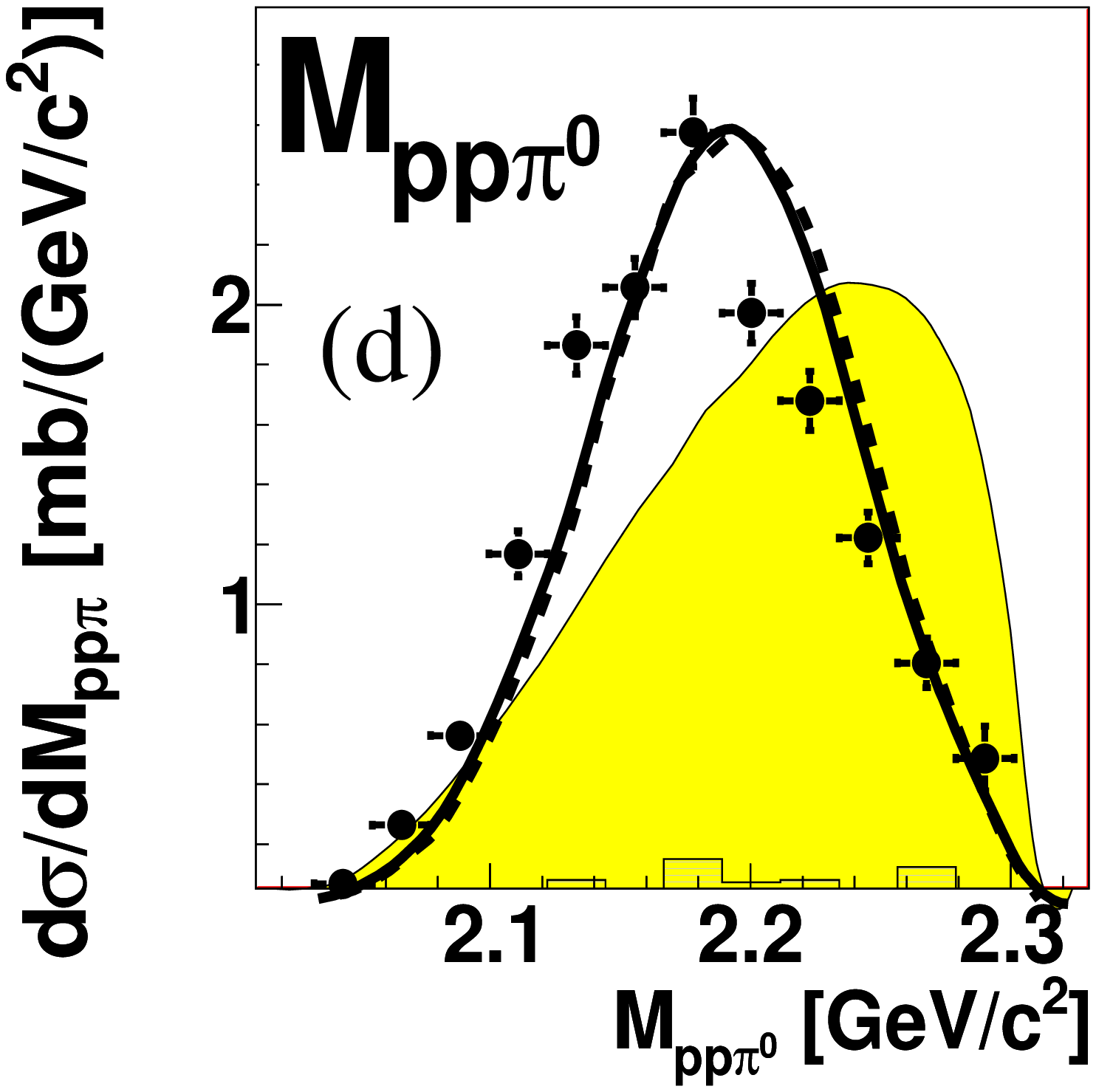}

\caption{(Color online) 
  Same as Fig. 3 but for the 
   distributions of the invariant masses $M_{pp\pi^0}$. 
}
\label{fig6}
\end{center}
\end{figure}



\begin{figure}
\begin{center}

\includegraphics[width=0.23\textwidth]{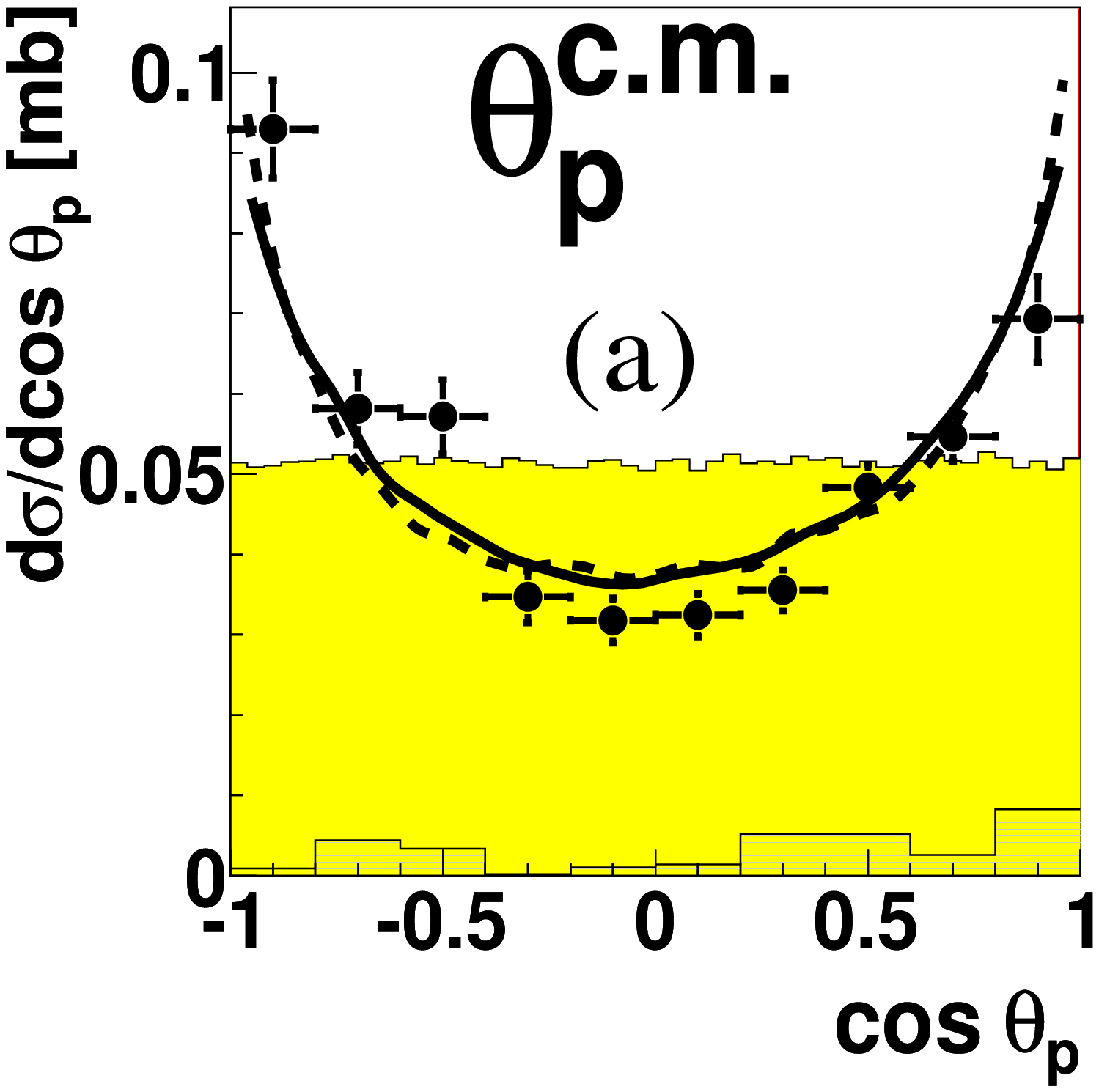}
\includegraphics[width=0.23\textwidth]{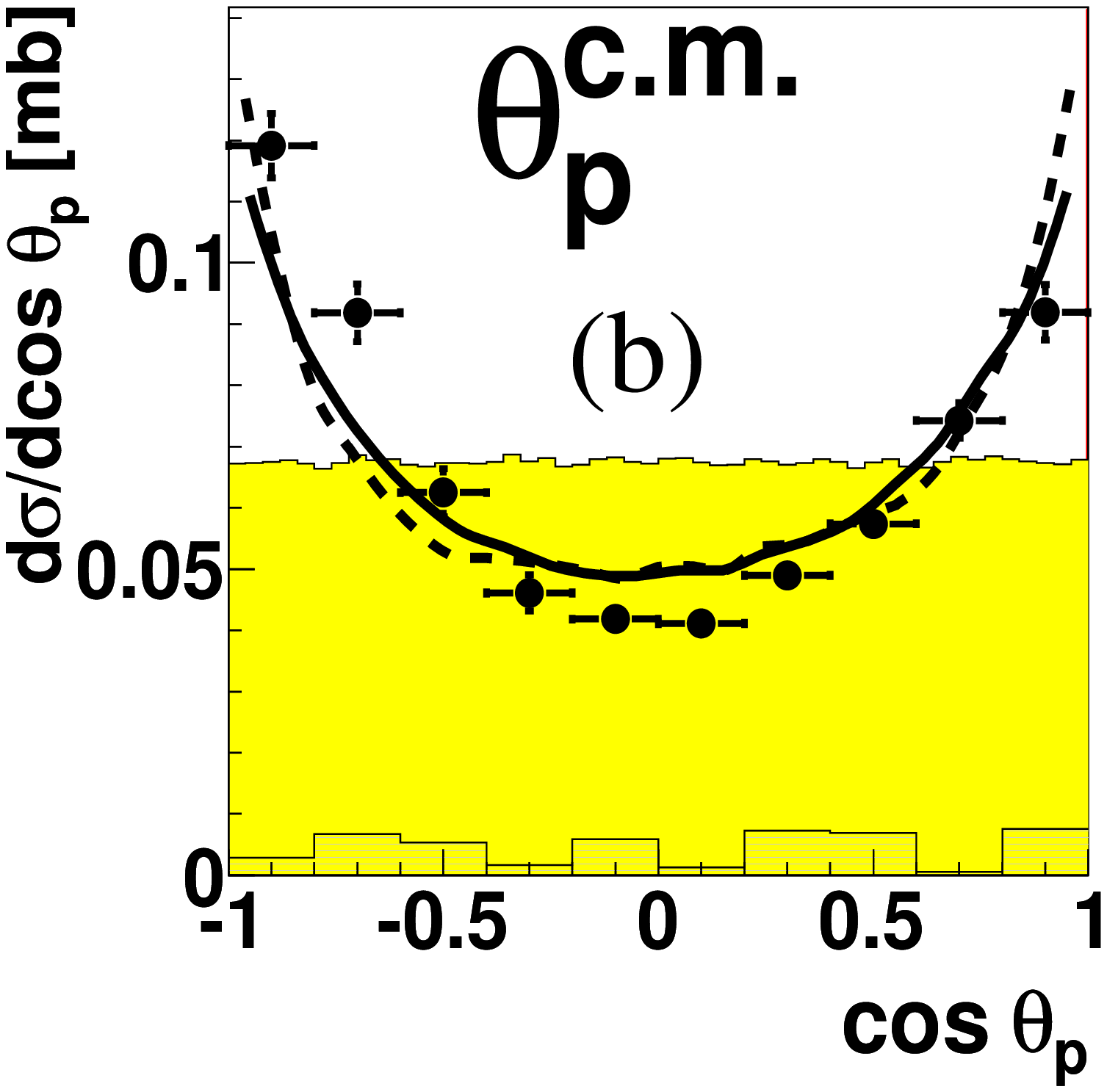}
\includegraphics[width=0.23\textwidth]{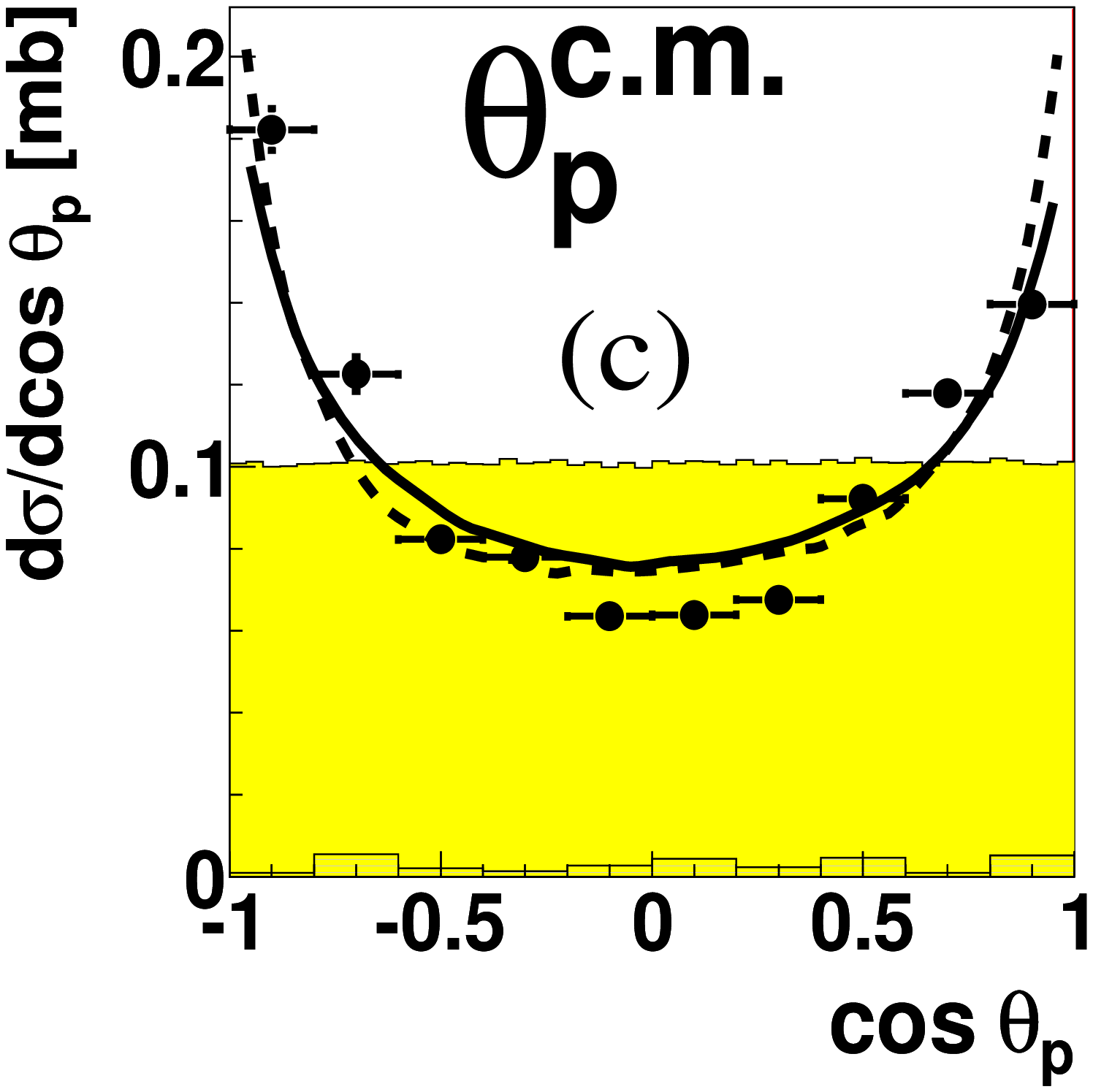}
\includegraphics[width=0.23\textwidth]{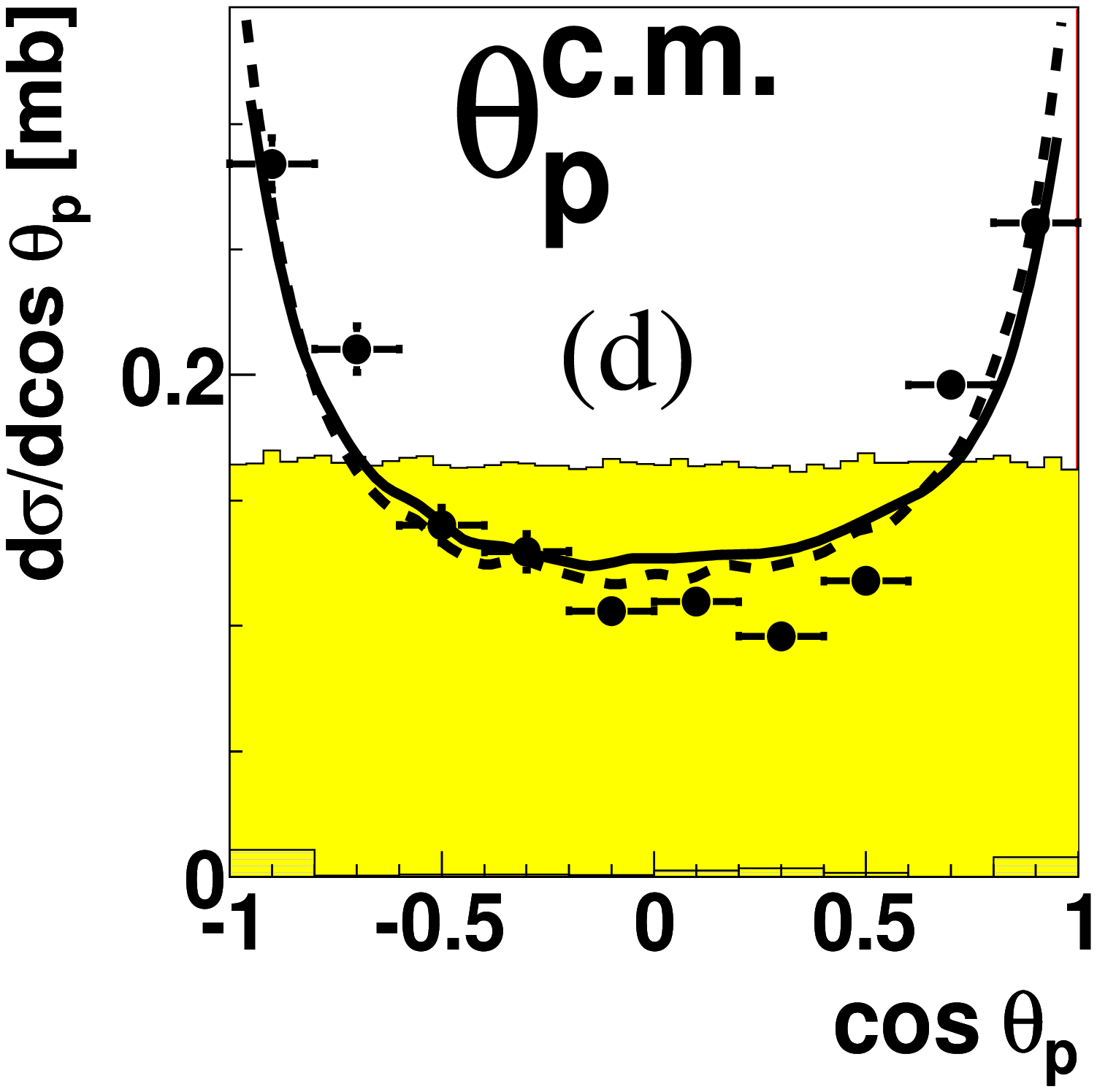}

\caption{(Color online) 
  Same as Fig. 3 but for the 
   distributions of the cms angle $\Theta_p$. 
}
\label{fig7}
\end{center}
\end{figure}

\begin{figure}
\begin{center}

\includegraphics[width=0.23\textwidth]{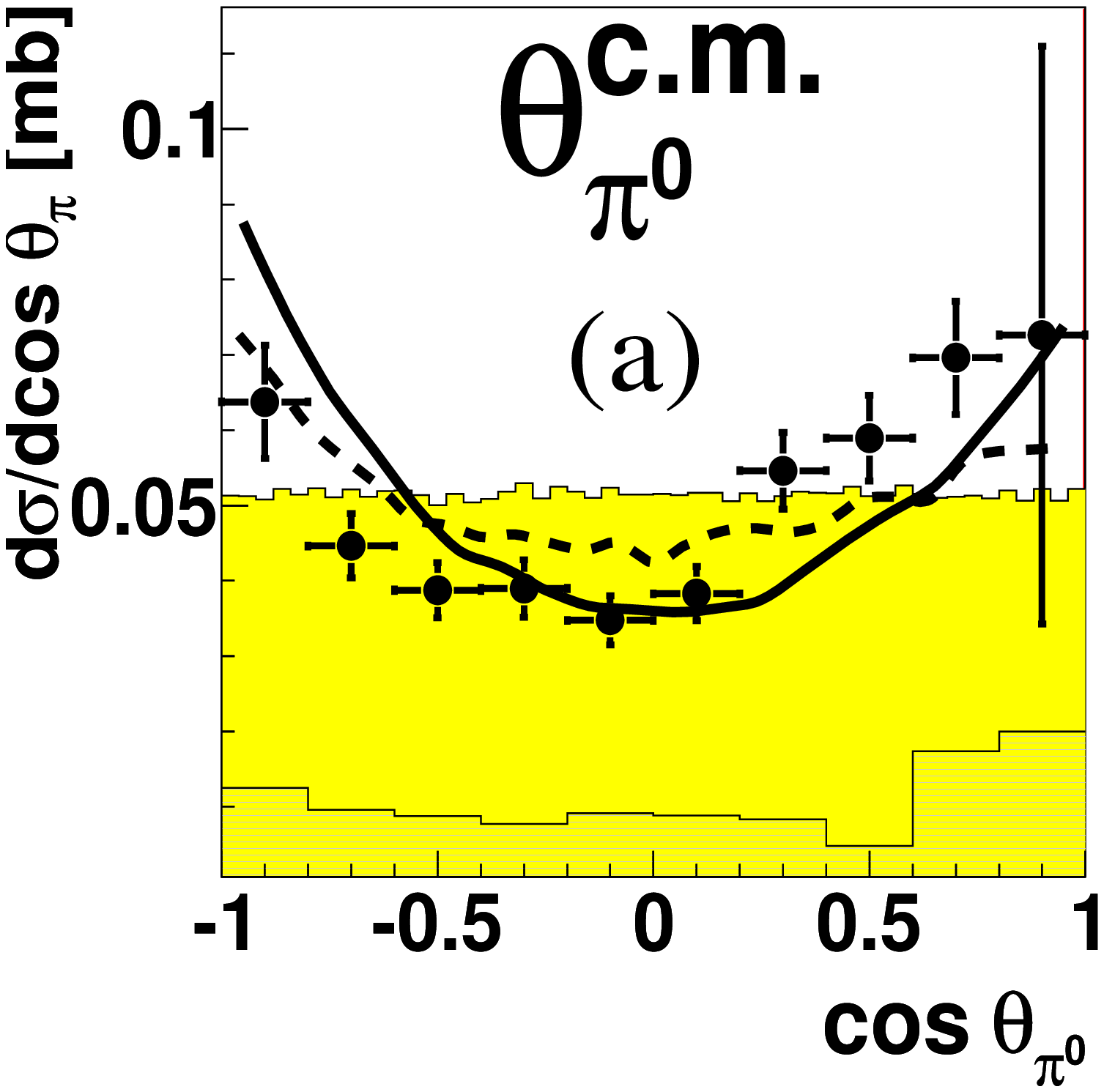}
\includegraphics[width=0.23\textwidth]{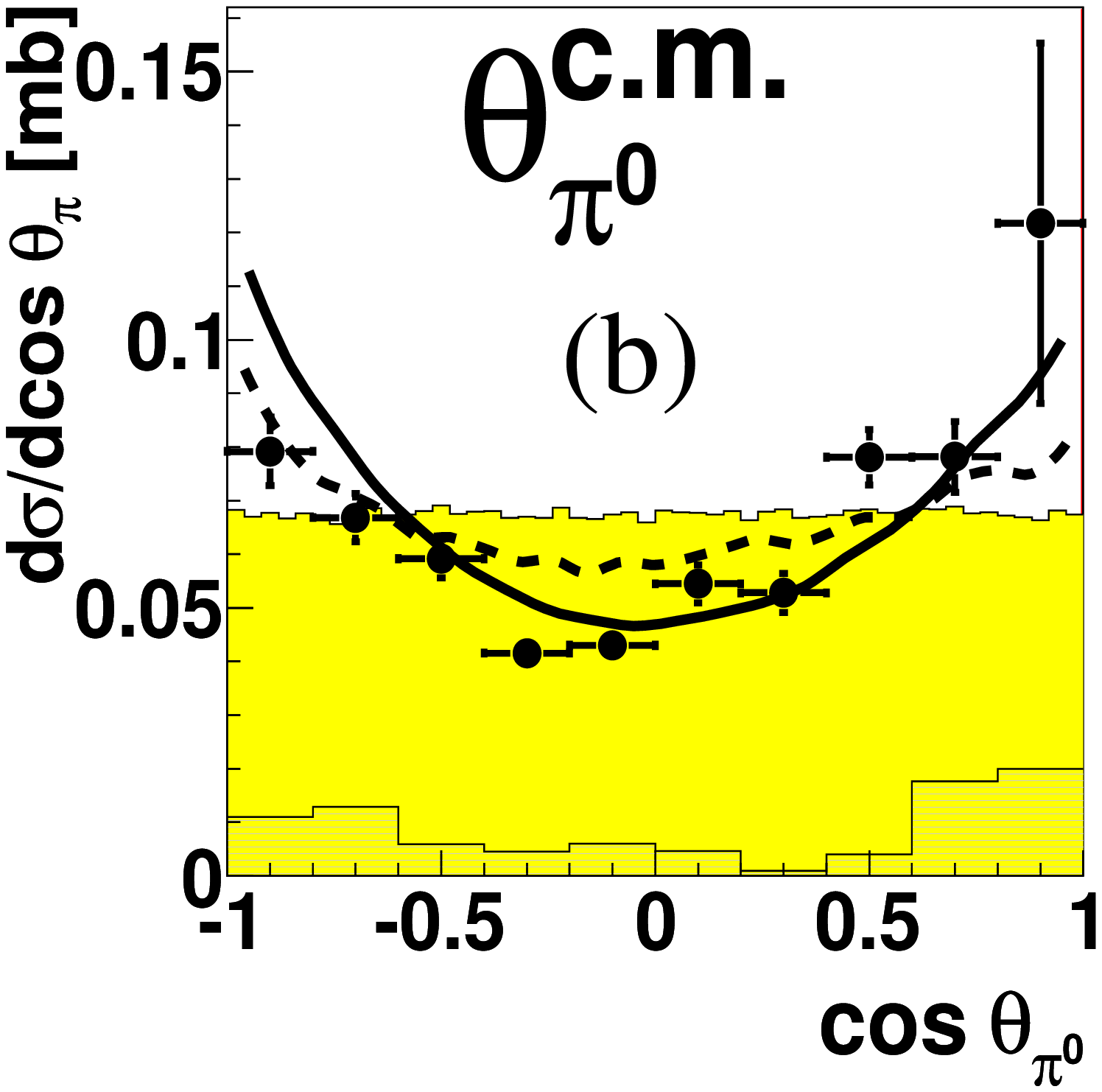}
\includegraphics[width=0.23\textwidth]{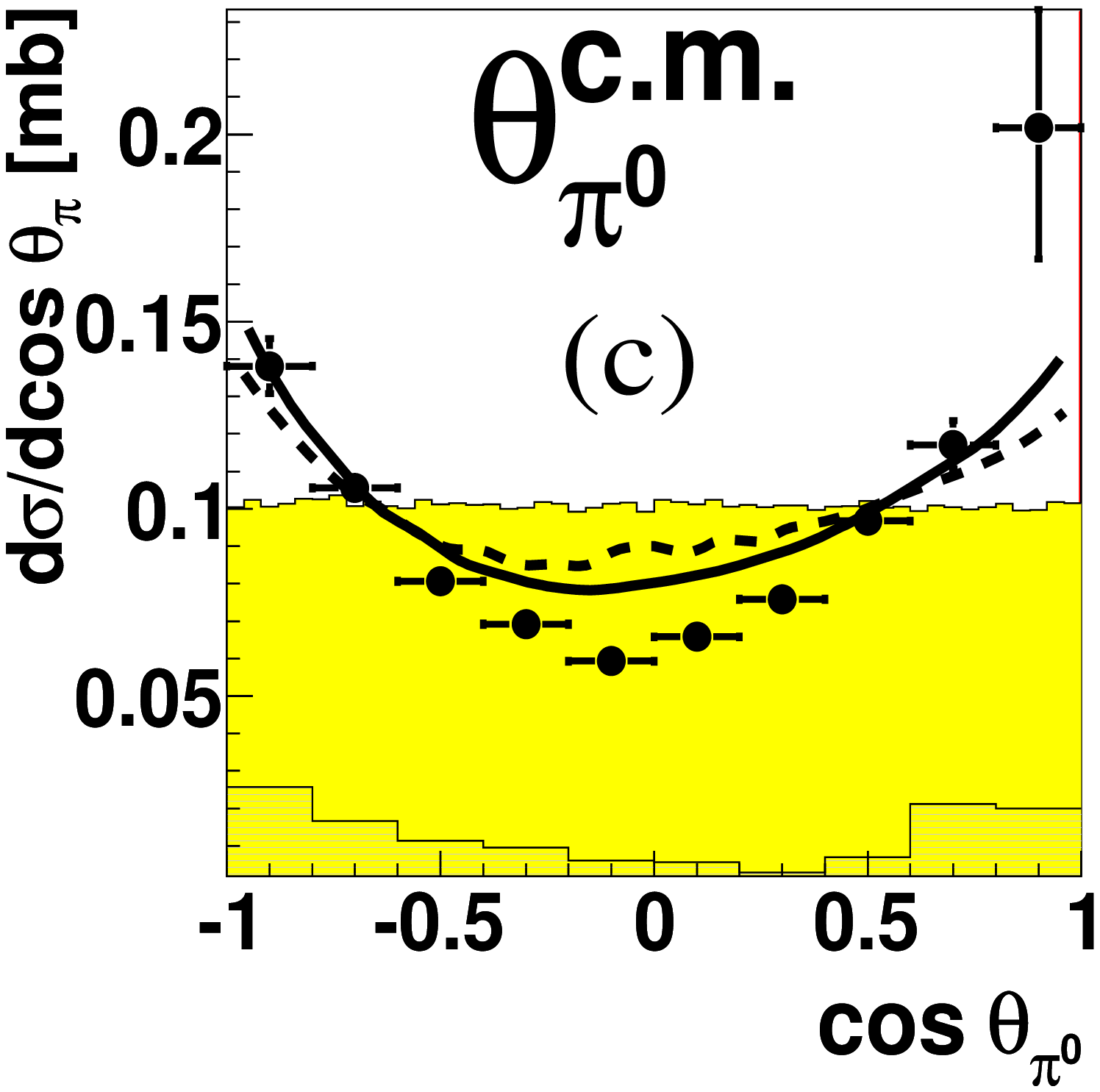}
\includegraphics[width=0.23\textwidth]{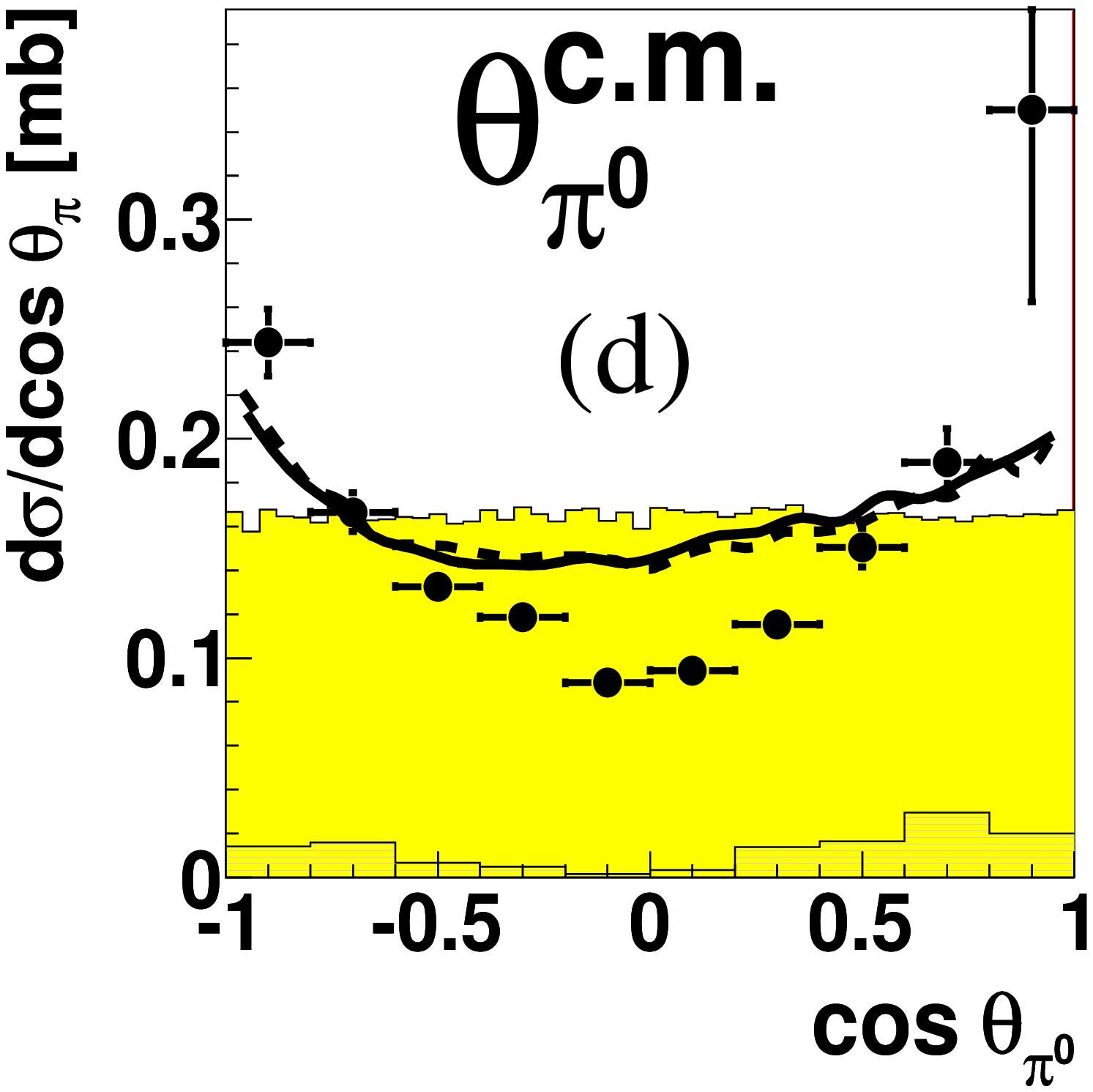}

\caption{(Color online) 
  Same as Fig. 3 but for the 
   distributions of the cms angle $\Theta_{\pi^0}$. 
}
\label{fig8}
\end{center}
\end{figure}

\begin{figure}
\begin{center}

\includegraphics[width=0.23\textwidth]{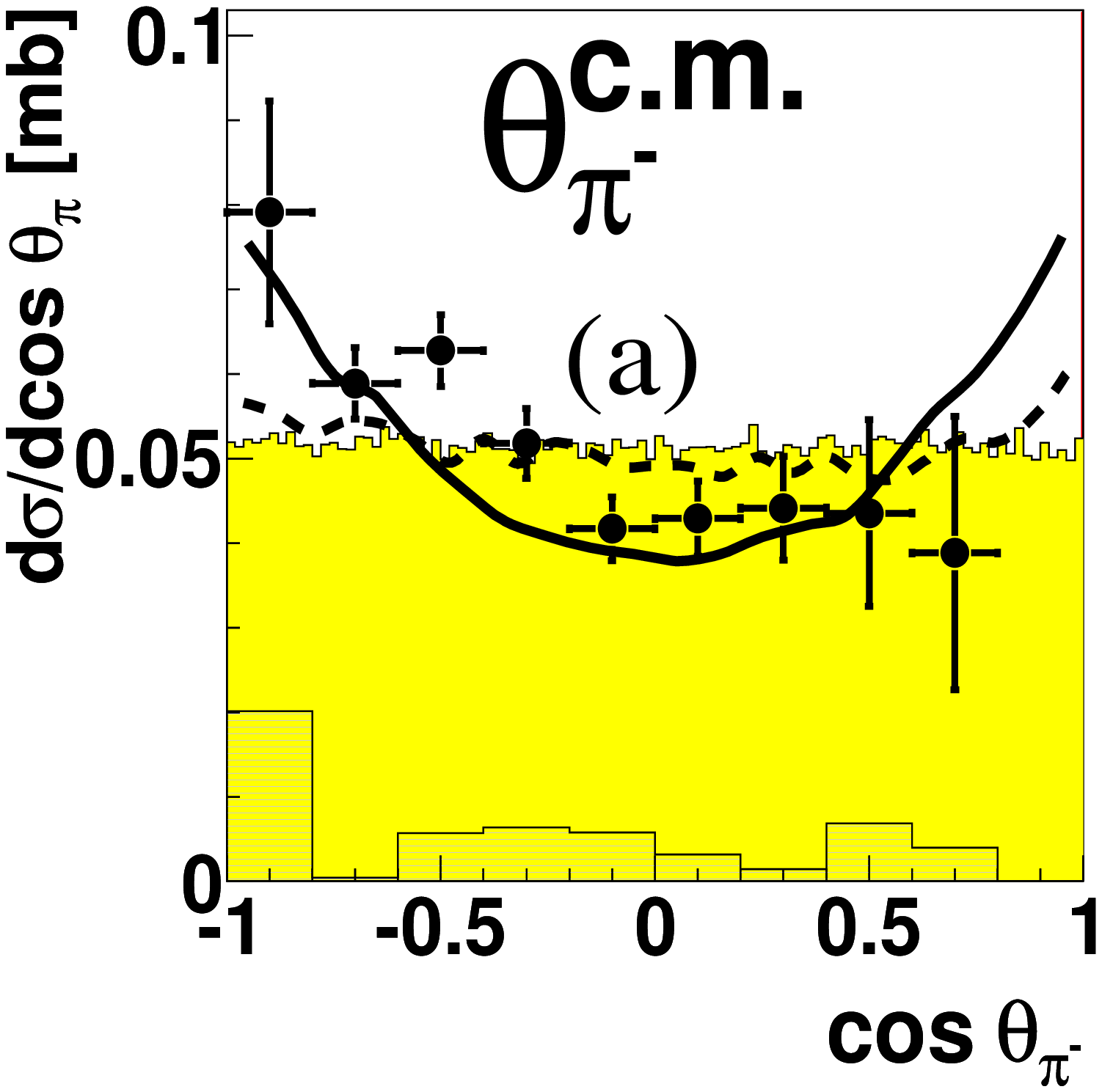}
\includegraphics[width=0.23\textwidth]{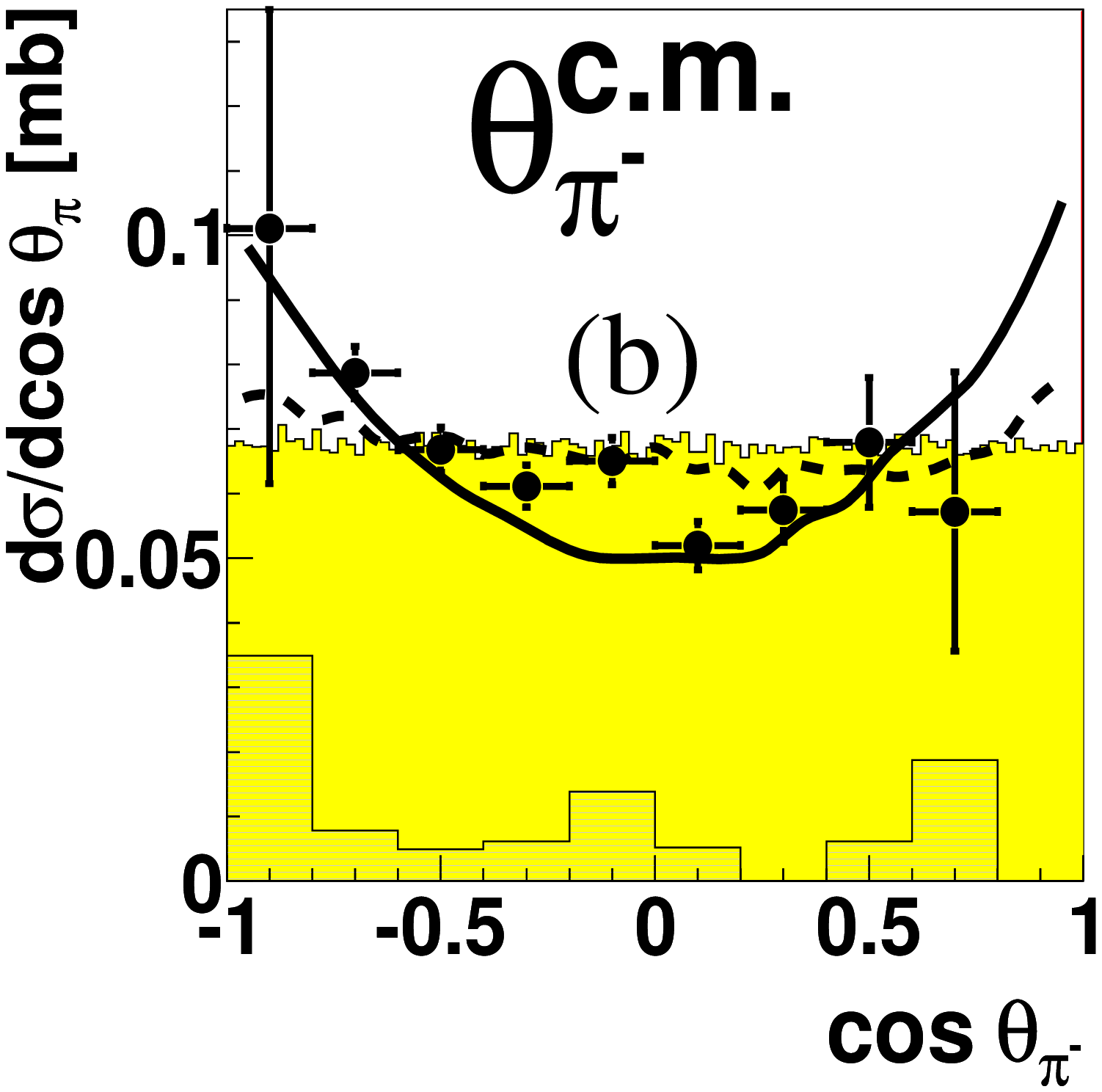}
\includegraphics[width=0.23\textwidth]{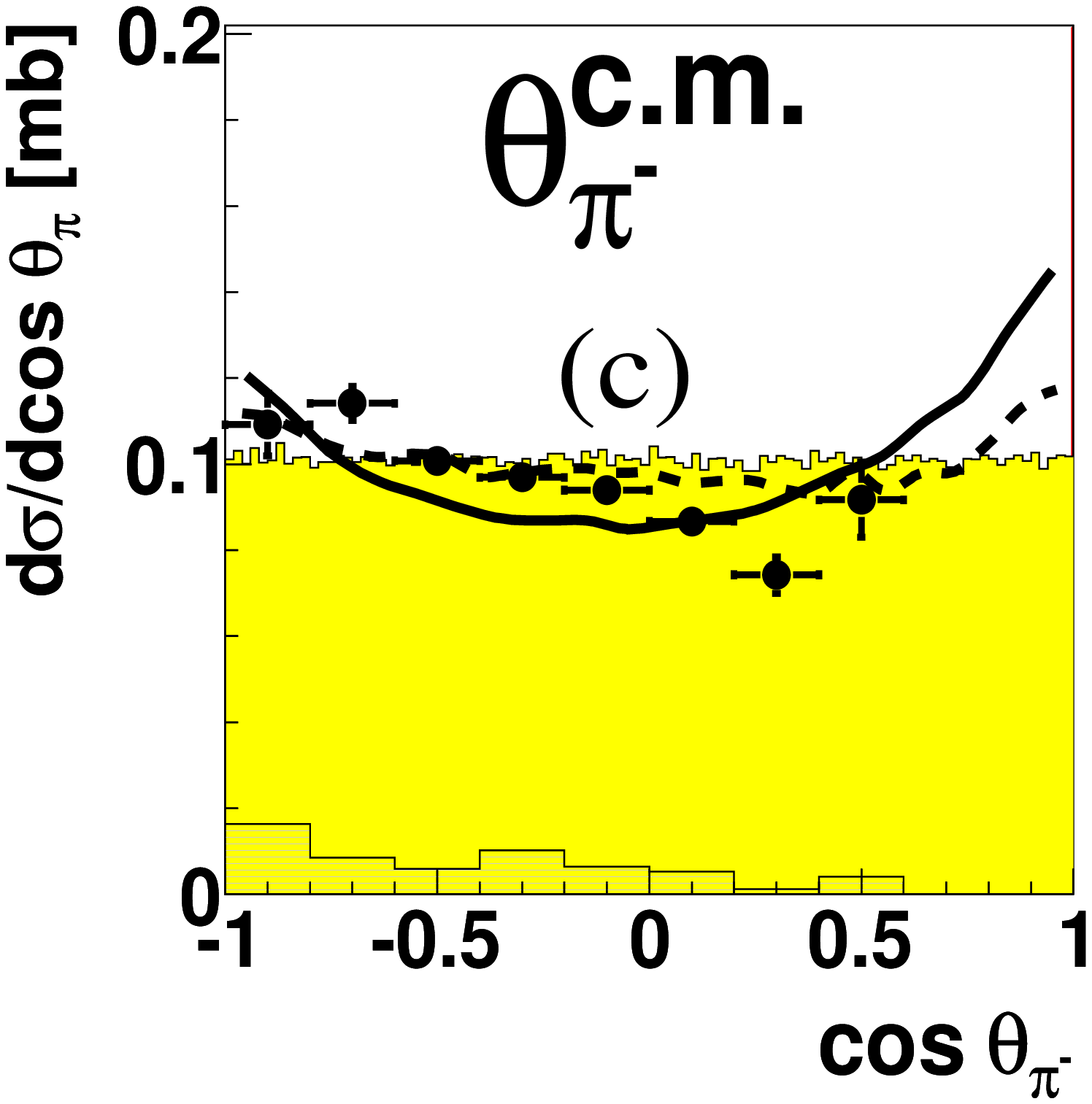}
\includegraphics[width=0.23\textwidth]{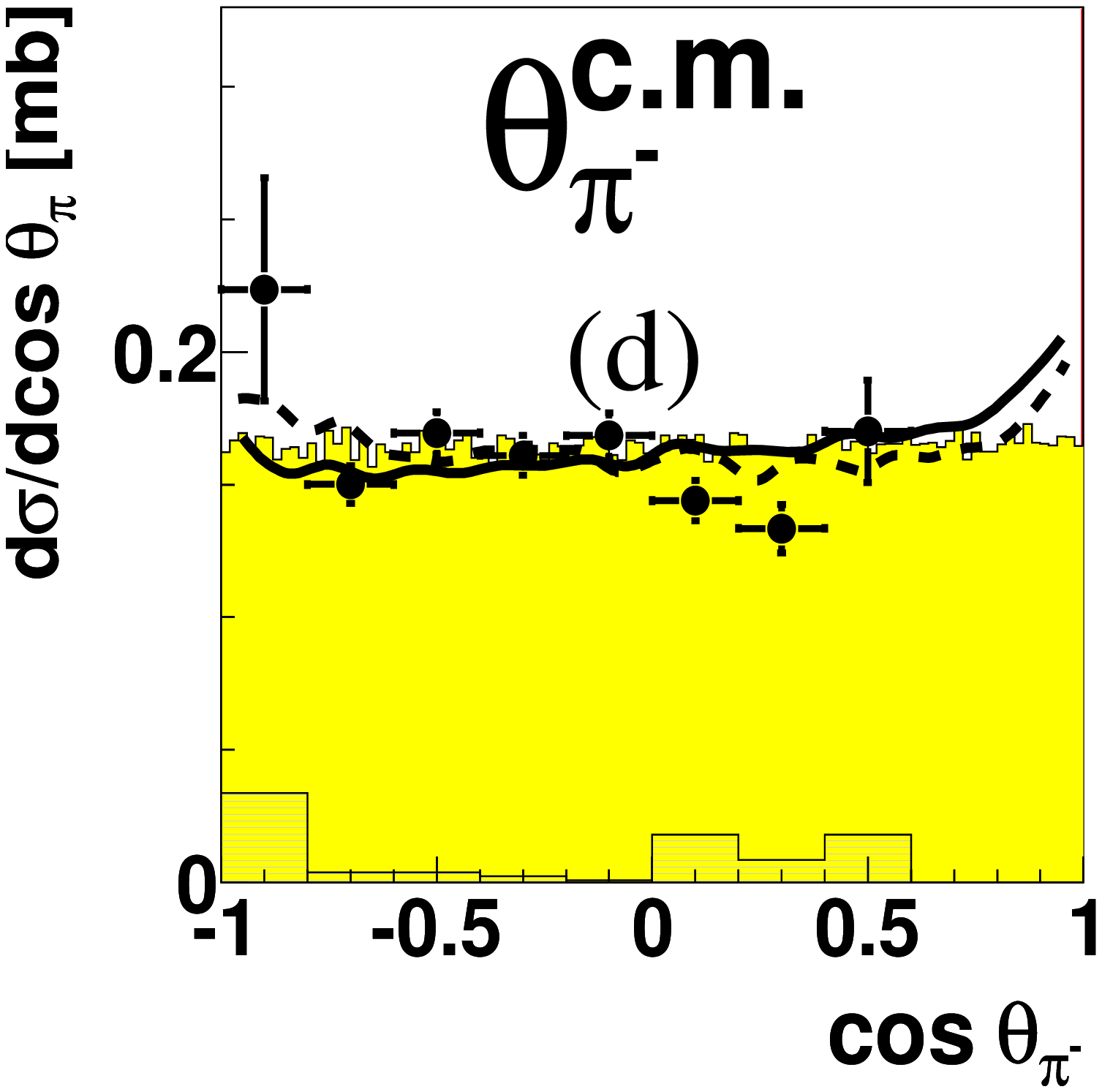}

\caption{(Color online) 
  Same as Fig. 3 but for the 
   distributions of the angle $\Theta_{\pi^-}$.
}
\label{fig9}
\end{center}
\end{figure}







In all cases we find an only gradual change in the shapes of the differential
distributions. At all energies the invariant mass distributions are
significantly different from pure phase space distributions (shaded areas in
Figs. 3 - 9). At the highest energy bin the observed invariant mass
distributions follow closely the shapes expected from the $\Delta\Delta$
process. This gets
particularly clear in the  $M_{p\pi^-}$ (see Fig.~4) and $M_{p\pi^0}$ (not
shown) spectra, where
pronounced peaks due to the $\Delta$ excitation develop -- compare corresponding
spectra in the $pp \to pp \pi^0\pi^0$ channel \cite{deldel}. Actually all
spectra are qualitatively similar in shape to those obtained in the $pp \to pp
\pi^0\pi^0$ channel with the exception of the $M_{\pi\pi}$ spectra
(Fig.~3). These observations are understandable by the fact that on the one
hand the $\Delta\Delta$ process is the leading process at high energies in both
channels, but on the other hand the $\pi\pi$ systems have different relative
angular momenta in these cases due to Bose symmetry. Whereas the isoscalar
$\pi^0\pi^0$ system is in relative $s$-wave, the isovector $\pi^0\pi^-$ system
has to be in relative $p$-wave. The $p$-wave condition favors large relative
momenta between the pions and hence causes a suppression of
intensity at low $\pi\pi$-masses and an enhancement at large masses compared
to phase space -- and that is what is indeed observed in the $M_{\pi^0\pi^-}$
spectra. 

From Fig.~5 we see that the observed $M_{pp}$ spectra exhibit distributions,
which are substantially narrower then the corresponding phase-space
distributions. Obviously large relative momenta between the two protons are
suppressed in the reaction of interest. Again the situation is very similar to
that in the $pp\pi^0\pi^0$ channel and may be traced to the dominant
$\Delta\Delta$ contribution. The modified Valencia calculations reproduce these
spectra very well (dashed curves in Fig.~5).

The $M_{pp\pi^0}$ spectra (Fig.~6) peak at $M = M_{\Delta} + M_p$ as expected
for a $pp\pi^0$ subsystem within the $\Delta\Delta$ excitation process.

The proton angular distributions exhibit a strongly anisotropic shape in
agreement with a peripheral reaction process (Fig.~7). Also the
pion angular distributions exhibit a pronounced anisotropy -- see Figs.~8 and
9. Both for protons and pions the anisotropy is significantly larger than
observed in the $pp\pi^0\pi^0$ channel. In the latter the two pions can be in
relative $s$-wave, whereas here in the $pp\pi^0\pi^-$ channel they have to be
in relative $p$-wave.  
Both the modified Valencia calculations (dashed curves in Figs.~3 - 9) and
those including the $d^*$ resonance (solid curves) provide very
similar shapes for the differential distributions in reasonable agreement
with the data. This similarity may appear surprising on a first glance and
hence needs some detailed consideration. 

First, the observed strongly anisotropic proton angular
distribution is very close to the one expected for a $J = 3$ resonance -- see
Ref. \cite{prl2011}. However, it is also equally well accounted for by
$t$-channel pion exchange, which produces a prominent U-shape at energies
far above the $\pi\pi$ threshold, see also Refs. \cite{deldel,TT}. 

Second, we expect a sizable effect from the dipole form factor at the
$\Delta\Delta$ vertex, which was introduced phenomenologically for the
description of the ABC effect, {\it i.e.} the low-mass enhancement in the
$M_{\pi^0\pi^0}$ distribution, in the $pn \to d\pi^0\pi^0$ reaction
\cite{prl2011}. Different from 
the bound nucleus case, where the relative momentum between the two $\Delta$s
is essentially made up by the relative momentum between the two emerging
pions, in the unbound case the relative $\Delta\Delta$ momentum is mainly
transferred to the two emerging nucleons -- the heavy partners of the $\Delta$
decays. Hence in the case of unbound nucleons in the final state we expect the
low-mass enhancement due to this form factor not to be in the $M_{\pi\pi}$
spectrum, but in the $M_{pp}$ spectrum. And this is also, what initially 
calculations with the inclusion of form-factor for the $d^*$ resonance
show. However, this effect is counterbalanced by the requirement that the two
protons have to be in relative $p$-wave, in order to build a $s$-channel
resonance with $J^P = 3^+$. In case of a
$d\pi^0\pi^0$ final state this spin-parity can be easily achieved by combining
the spin 1 of the deuteron with the p-wave decays of the two $\Delta$ states
into the $N\pi$ system such that in total we have a $\pi^0\pi^0$ system in
relative $s$-wave, which again is in relative $d$-wave to the deuteron. In
case of the 
$pp\pi^0\pi^-$ channel we have an isovector $\pi^0\pi^-$ system, which by Bose
symmetry need to be in relative $p$-wave. To fulfill the required spin-parity,
the $pp$ system can no longer be in relative $s$-wave, but needs to be at
least in a relative $^3P_2$ state.

That way, {\it i.e.} by inclusion of the $d^*$ resonance, we obtain a
description for both integral (solid curve in Fig.~2) and differential cross
sections (solid curves in Figs.~3 - 9), which is comparable in quality to what
was achieved for the description of the the purely isovector channels
$pp\pi^0\pi^0$ and $nn\pi^+\pi^+$. 


Concerning the $\Delta\Delta$ vertex
form factor, which was introduced for the phenomenological description of the
ABC effect in the $pn \to d\pi^0\pi^0$ reaction, we would like to mention an
alternative ansatz proposed recently by Platinova and Kukulin
\cite{kuk}. They 
assume the $d^*$ resonance not only to decay into the $d\pi^0\pi^0$ channel
via the route $d^* \to \Delta^+\Delta^0 \to d\pi^0\pi^0$, but also via the
route $d^*
\to d\sigma \to d\pi^0\pi^0$. Since $\sigma$ is a spin zero object, it has to
be in relative $d$-wave to the deuteron in this decay process, in order to
satisfy the resonance condition of $J^P~=~3^+$. In consequence the available
momentum in the decay process is concentrated in the relative motion between
$d$ and $\sigma$ leaving only small relative momenta between the two emerging
pions. Therefore the $M_{\pi^0\pi^0}$ distribution is expected to be peaked at
low masses. {\it I.e.}, the low-mass enhancement (ABC effect) in this model is
made by the $d\sigma$ decay branch and not by a form factor as introduced in
Ref. \cite{prl2011}.  The enhancement in this model is further increased by
interference of the $d\sigma$ decay amplitude with the
decay amplitude via the $\Delta^+\Delta^0$ system. Applying this scenario to
the $pp\pi^0\pi^-$ channel we have in this case no decay branch via the
isoscalar $\sigma$ 
configuration, since the $\pi^0\pi^-$ pair is purely isovector. Hence the
$d^*$ decay into this channel proceeds solely via the $\Delta^+\Delta^0$
system and does not exhibit any low-mass enhancement (ABC effect) -- neither
in the $M_{\pi^0\pi^-}$ nor in the $M_{pp}$ system. This situation corresponds
just to a $d^*$ calculation without form factor at the $\Delta^+\Delta^0$
vertex. Since then the $p$-wave condition for the $pp$
subsystem is no longer counterbalanced by the effect of the form factor, the
calculated $M_{pp}$ distribution gets wider and close to phase-space worsening
thus somewhat the agreement with the data.



 

\section{Summary and Outlook}

The first exclusive and kinematically complete $pn \to pp\pi^0\pi^-$
measurements of solid statistics have been carried out in quasifree kinematics
with a proton beam hitting a deuterium target. Utilizing the
nucleons' Fermi motion in the deuterium target an energy region of
2.35 GeV $< \sqrt s <$ 2.46 GeV could be covered corresponding to an incident
lab energy range of (1.07 - 1.36) GeV. This energy region also covers the region
of the ABC effect and its associated narrow resonance structure around 2.37
GeV. No evidence for a low-mass enhancement (ABC effect) is found in the
data for the $\pi^0\pi^-$-invariant mass distribution. Its absence is easily 
understood from the fact that the isovector $\pi^0\pi^-$ pair has to be in
relative $p$-wave and -- even more important -- that in this case of unbound
nucleons the form factor introduced for the description of
the ABC effect in the $d\pi\pi$ channel causes a low-mass enhancement in
$M_{pp}$ and not in $M_{\pi\pi}$. In the latter, however, the impact of the form
factor is counterbalanced by the condition that the two protons have to be
in relative $p$ wave, in order to reach the  $J^P = 3^+$ requirement for the
resonance. 

The differential data are reasonably well accounted for by conventional
$t$-channel calculations with the modified Valencia model
\cite{luis,deldel,nnpipi}. These calculations also give a good description of
the total cross section at the highest measured energies. However, at lower
energies these calculations lack up to at least a factor of four in cross
section. Since such a big failure has not been observed in $pp$-induced
reaction channels and since it concerns the low energy region, where no
$t$-channel resonance processes are known to contribute, it has to be
ascribed to an unconventional isoscalar process. One such
process is the excitation of the $d^*$ resonance. Its inclusion in the model
description for the $pn \to pp\pi^0\pi^-$ reaction leads to a much improved 
understanding of both differential and total cross section data. The necessary
peak cross section of about 100~$\mu$b for the $d^*$ contribution agrees very
well with expectations. 

After the experimental evidences found in the $d\pi^0\pi^0$
and $d\pi^+\pi^-$ channels, the $ pp\pi^0\pi^-$ channel is now the third
channel, which is consistent with the $d^*$ hypothesis. If true, then this
resonance should also been sensed in the $pn \to pn\pi^0\pi^0$ reaction and --
most importantly -- in $pn$ scattering, the {\it experimentum crucis}. Data
for these reactions have been taken already by the WASA collaboration. Their
analysis is in progress.


\section{Acknowledgments}

We acknowledge valuable discussions with 
C. Hanhart, 
V. Kukulin, E. Oset,
A. Sibirtsev and C. Wilkin on this issue. We are particularly indebted to
L. Alvarez-Ruso for using his code.  
This work has been supported by BMBF
(06TU9193), Forschungszentrum J\"ulich (COSY-FFE), the Polish National Science
Centre and the Foundation for Polish Science.

\end{document}